\documentstyle[aps,epsfig]{revtex}
\newcommand{\n}{\nonumber}
\begin{document}
\title{``Sum over Surfaces'' form of Loop Quantum Gravity}
\author{Michael P. Reisenberger\footnote{miguel@fisica.edu.uy}}
\address{Instituto de F\'{\i}sica, Universidad de la Rep\'{u}blica,
        Tristan Narvaj\'a 1674, 11200 Montevideo, Uruguay;\\
   Erwin Schr\"odinger Institute, A-1090 Vienna, Europe.}
\author{Carlo Rovelli\footnote{E-mail: rovelli@pitt.edu}}
\address{Department of Physics and Astronomy, 
	University of Pittsburgh, Pgh Pa 15260, USA;\\
	Center for Gravity and Geometry, Penn State University,
	State College, Pa 16801, USA;\\
   Erwin Schr\"odinger Institute, A-1090 Vienna, Europe.}
\maketitle
\date{\today}

\vspace{.2cm} \begin{abstract} We derive a {\em spacetime\/} formulation of
quantum general relativity from (hamiltonian) loop quantum gravity.  In
particular, we study the quantum propagator that evolves the 3-geometry in
proper time. We show that the perturbation expansion of this operator is
finite and computable order by order.  By giving a graphical
representation {\em \`a la\/} Feynman of this expansion, we find that the
theory can be expressed as a sum over topologically inequivalent
(branched, colored) 2d surfaces in 4d.  The contribution of one surface to
the sum is given by the product of one factor per branching point of the
surface.  Therefore branching points play the role of elementary vertices
of the theory.  Their value is determined by the matrix elements of the
hamiltonian constraint, which are known. 

The formulation we obtain can be viewed as a continuum version of 
Reisenberger's simplicial quantum gravity.  Also, it has the same 
structure as the Ooguri-Crane-Yetter 4d topological field theory, with a
few key differences that illuminate the relation between quantum gravity
and TQFT.  Finally, we suggests that certain new terms should be added to
the hamiltonian constraint in order to implement a ``crossing'' symmetry
related to 4d diffeomorphism invariance. 

gr-qc/9612035 \hskip 9cm May 1997

\end{abstract}
\vskip1cm

\section{Introduction}

An old dream in quantum gravity \cite{Isham96} is to define a 
manifestly spacetime-covariant Feynman-style ``sum over 
trajectories'' \cite{hh}, sufficiently well defined to yield finite
results order by order in some expansion.  The hamiltonian theory has 
obtained encouraging successes in recent years, but it suffers 
for the well-known lack of transparency of the frozen time formalism, 
for the difficulty of writing physical 
observables and for operator ordering ambiguities. These problems 
are related to the lack of manifest 4d covariance.  Here, we derive a 
covariant spacetime formalism {\it from\/} the hamiltonian theory.  This 
is of course the path followed by Feynman to introduce his sum over 
trajectories in the first place \cite{fey}.  What we obtain is 
surprising: we obtain a formulation of quantum gravity as a sum over 
surfaces in spacetime.  The surfaces capture the gravitational degrees 
of freedom.  The formulation is ``topological'' in the sense that one 
must sum over topologically inequivalent surfaces only, and the 
contribution of each surface depends on its 
topology only\footnote{More precisely: on the diff-invariant properties 
of the surface.}.  This contribution is given by the product of 
elementary ``vertices'', namely points where the surface branches.  The 
sum turns out to be finite and explicitly computable order by 
order.  The main result of this paper is the construction of this 
finite ``sum over surfaces'' formulation of quantum gravity.

Let us sketch here the lines of the construction. Given gravitational data
on a spacelike hyper-surface $\Sigma_i$, the three-geometry on a surface
$\Sigma_f$ at a proper time $T$ in the future of $\Sigma_i$ (as measured 
along geodesics initially at rest on $\Sigma_i$), is uniquely determined 
in the classical theory.  It is then natural to study the corresponding 
evolution  operator $U(T)$, that propagates states from $\Sigma_i$ to 
$\Sigma_f$ in the quantum theory.  This operator, first considered by 
Teitelboim \cite{claudio}, codes the dynamics of the quantum 
gravitational field, and is analogous to the Feynman-Nambu proper time 
propagator \cite{fn} for a relativistic particle.  Here, we construct the 
operator $U(T)$ in quantum GR, and we expand it in powers of $T$.  We 
obtain a remarkable result: the expansion is finite order by order.  
This is our first result.

Next, we construct a graphical representation of the expansion.  This is 
obtained by observing that topologically inequivalent colored 2d surfaces 
$\sigma$ in spacetime provide a natural bookkeeping device for the terms of 
the expansion.  We obtain an expression for $U(T)$ as a sum of terms 
labeled by surfaces $\sigma$ bounded by initial and final states.  A 
surface $\sigma$ consists of simple components (2-manifolds), or ``faces'', 
that carry a positive integer or color.  Faces meet on 1d ``edges'', 
colored as well.  Edges, in turn, meet at branching points, denoted 
``vertices''.    The weight of each surface in the sum is a product of 
factors associated to its vertices.  The value of a vertex is determined by 
the hamiltonian constraint, and is given by a simple function (involving 
Wigner $3n$-$j$ symbols) of the colors of the adjacent faces and edges.  
This ``sum over surfaces'' version of the dynamics of quantum general 
relativity is our second result.

The construction allows us to consider transformation properties of the 
hamiltonian constraint under 4d diffeomorphisms (diff) 
in a manifestly covariant way.  This analysis suggests the 
addition of certain new terms to the constraint, corresponding to an 
alternative operator ordering, which implements a ``crossing'' symmetry at 
the vertices, with a nice geometrical appeal.  Thus, 4d diff invariance may 
be a key for reducing the present ambiguity in the operator ordering. 
Furthermore, the new terms seem to prevent some potential problems  
with locality pointed out by Smolin \cite{lee}.  The introduction of these 
new terms in the quantum hamiltonian is our third result.

The idea that one could express the dynamics of quantum gravity in terms
of a sum over surfaces has been advocated in the past, particularly by
Baez \cite{baez} and Reisenberger \cite{mike2}.  On the lattice, a sum
over surfaces was recently developed by Reisenberger \cite{mike}, and the
lattice construction has guided us for the continuum case studied here. 
It is important to emphasize, however, that the present construction is
entirely derived from the canonical quantum theory in the continuum. 

The sum over surfaces we obtain has striking similarities with topological
quantum field theory (TQFT). More precisely, it has the same kinematic as
the Ooguri-Crane-Yetter model \cite{ooguri,CY}, a 4d TQFT which extends 
the Ponzano-Regge-Turaev-Viro 3d TQFT \cite{PR,TV,pri} to four dimensions.  
Essentially, the difference is given just by the weight of the
vertices.  In Appendix B we discuss similarities and differences between 
the two theories. The discussion, we believe, sheds much light over the 
tantalizing issue \cite{jmp} of the relation between 
finite-number-of-degrees-of-freedom TQFT and quantum gravity.  In
particular, we argue that a diff-invariant quantum field theory with an
{\it infinite\/} number of local (but non-localized) degrees of freedom
--such as quantum general relativity-- can be obtained by having a sum
over arbitrarily fine triangulations, instead of triangulation independence,
as in combinatorial TQFTs. 

On the other hand, the sum over surfaces we obtain can be viewed as (a
first step towards) a concrete implementation of Hawking's sum over
4-geometries \cite{hh}. In fact, the surfaces over which we sum have an
immediate interpretation as ``quantum'' 4-geometries, as we will
illustrate.  This fact should make the the general techniques of covariant
generalized quantum mechanics \cite{hartle} available to quantum gravity,
potentially simplifying the difficulties with physical observables of 
the hamiltonian formalism. 

The basis of our construction is loop quantum gravity 
\cite{loops,ashtekar,loops2}. 
The finiteness of the sum-over-surfaces and the picture of a ``discrete 
4-geometry'' that emerges from this work are related to the fact that 
geometrical operators have discrete spectra.  The discreteness of the 
spectra of area and volume --and the ``quantized'' structure of space 
that these spectra suggests-- is a central result in loop quantum 
gravity, first obtained by Rovelli and Smolin in \cite{discr}, and later 
confirmed and clarified by a number of authors 
\cite{disc2,loll,dpr,fl,volume}.   The main ingredient of our 
construction is the quantum hamiltonian constraint 
\cite{loops,hamiltonian}. 
In particular, Thiemann's version of the hamiltonian constraint 
\cite{Thiemann} and some variants of it play an essential role here.  
Matrix elements of this operator have been computed explicitly in 
\cite{rrr}, using the methods developed in \cite{dpr}.\footnote{
Note added to the revised version: An extremely interesting 
sequence of developments has appeared after the first version  
of this work.  Smolin and Markopoulou \cite{sm} have attempted
to incorporate the causal structure of the lorentzian theory 
directly into the surfaces' kinematic. Baez \cite{b} has explored 
the general structure of the theories of the kind introduced here,
namely defined by a sum over surfaces with weights at the vertices.  
Markopolou \cite{fotini} and Baez  \cite{b2} have illuminated the 
nature of the vertices by consider evolution in the context of  
triangulated spacetimes.  In a triagulated manifold "time evolution" can 
be decomposed into a finite number of elementary moves, corresponding to 
the possible ways of adding a (spacetime) (3+1)-simplex to a (space) 3-d 
boundary. When viewed in the {\it dual/} of the triangulation, these 
elementary moves correspond to the (crossing symmetric) vertices generated
by the hamiltonian constraint operator and studied in this paper.
We find this convergence fascinating.  Finally, Crane \cite{crane2} 
and Baez \cite{b2} have explored two possible ways of obtaining  
quantum GR by modifying the Ooguri-Crane-Yetter state sum. (On this,
see also \cite{mike2}.)  We think that deriving the sum over surfaces 
formulation of quantum GR directly from the lagrangian would be a major 
step ahead and would shed light on the problem of the integration over 
the lapse.}

In section II, we summarize the basics of nonperturbative loop quantum
gravity.  In section III we define the proper time propagator and its
expansion.  In section IV we show that the proper time propagator can be
expressed as a sum over surfaces.  In section V we discuss crossing
symmetry and the new terms of the hamiltonian constraint.  In section VI
we summarize and comment our results. Appendix A 
is a brief glossary of some geometrical term employed.  Appendix B
contains the comparison with TQFT.  In Appendix C we give an example of 
3d diff-invariant scalar product. A preliminary version of this
work has appeared in \cite{carlo}.

\section{Canonical Loop Quantum Gravity}

\subsection{Kinematic}

We start with nonperturbative canonical quantum gravity in the loop 
representation \cite{loops}.  The Hilbert space ${\cal H}$ of the 
theory is spanned by the basis $|S\rangle$, where $S$ is a spin 
network \cite{spin}.  A spin network is a colored graph $\Gamma$ 
embedded in a (fixed) three dimensional compact manifold 
$\Sigma$.\footnote{We recall the definition of coloring of a spin 
network \cite{dpr}.  Each node of the graph with valence higher than 
three (more than three adjacent links) is arbitrarily expanded in a 
tree-like trivalent subgraph.  The internal links of the subgraph are 
denoted virtual links.  The coloring of the graph is an assignment of 
a positive integer to each real or virtual link -- in such a way that 
at every trivalent node the sum of the three colors is even and none 
of the colors is larger than the sum of the other two.  The set of the 
colorings of the virtual links of a node is also called coloring of 
the node.  A coloring can be thought as an assignment of an irreducible 
$SU(2)$ representation to each link and of an invariant coupling tensor 
to each node.} For a fixed choice of a tree-like expansion at the nodes, 
these states are orthonormal \cite{dpr,sn,inverse} 
(see \cite{dpr} for details on their normalization).
\begin{equation} 
\langle S^\prime|S\rangle= \delta_{SS^\prime}; 
\label{ip} 
\end{equation} 
the matrix elements of the change of basis between different nodes' 
expansion can be derived from equation (\ref{recoupling}).
An equivalent construction of this Hilbert space can be obtained 
in terms of functions over (generalized) connections \cite{ashtekar}. 
For details on the equivalence between the two formalism, see 
\cite{rob}.

The dynamics of quantum general relativity is governed by two operators 
in $\cal H$: the Diff constraint operator $C[\vec N]$ and the hamiltonian 
constraint $C_L[N]$.  Let us examine them.

\subsection{The Diff constraint and its solutions}
 
For every diffeomorphism $f:\Sigma \rightarrow \Sigma$ (in
$Dif\! f_0$, the component of the diffeomorphism group connected to 
the identity), let $D[f]$ be the operator in $\cal H$ giving the 
natural action $f:  S\longmapsto f\cdot S$ of the diffeomorphism on the spin
network states.  Namely
 \begin{equation} 
D[f]\ |S\rangle=|f\cdot S\rangle. 
 \end{equation} 
 For every vector field $\vec N$ on $\Sigma$ that generates a one
parameter family $f_t$ of diffeomorphisms (by $df_t/dt = \vec
N$, $f_0$=identity)\footnote{We put an arrow over vectors  
($\vec N$); but not over spatial coordinates ($x$) or diffeomorphisms 
($f$).}  the Diff constraint is defined by\footnote{Rigorously speaking, 
$C[\vec N]$ is not well defined on $\cal H$. This is due to funny 
(kinematical) inner 
product (\ref{ip}), in terms of which the action of the 
diffeomorphism group is not strongly continuous. This fact does not 
disturb the construction of the theory, because the only role played 
by $C[\vec N]$ is to implement invariance under the finite 
transformations it generates. These are well defined 
\cite{ashtekar3}. Here, it is useful to consider $C[\vec N]$  
as well, because it plays a role in the formal manipulations 
below.} 
 \begin{equation} 
C[\vec N]= \left. -i {d\over dt}\ D[f_t]\right|_{t=0}; 
 \end{equation} 
and corresponds to the classical diffeomorphism constraint smeared 
with Shift function $\vec N$. The space ${\cal H}_{diff}$ of the 
solutions of the diffeomorphism constraints is defined as ${\cal 
H}_{diff}={{\cal H}\over Dif\!f_0}$. It is spanned by a basis 
$|s\rangle$, where $s$ is an s-knot, namely an equivalence class of 
spin networks under diffeomorphisms, which define the linear structure of 
${\cal H}_{diff}$.  One can define the scalar 
product in ${\cal H}_{diff}$ by an integration \cite{gi,higu} 
over $Dif\!f_0$.  If $S\in s$ and $S^\prime\in s^\prime$,
 \begin{equation}
 \langle s|s^\prime \rangle = {\cal N} \int_{Dif\!f_0}[df]\ \langle 
f\cdot
 S|S^\prime\rangle.
\label{product}
 \end{equation}
 $\cal N$ is a normalization factor. Equation (\ref{product}) is
meaningful because the integrand vanishes over most of the 
integration space (because two spin network states are orthogonal 
unless they have the same graph) and is constant on a discrete 
number of regions whose volume is normalized to one by $\cal N$.  
Thus we have
 \begin{equation}
 \langle s|s^\prime \rangle = \sum_\rho \langle \rho 
S_i|S^\prime\rangle, 
\label{product2}
 \end{equation}
 where the sum is over the (discrete) automorphisms $\rho$ that 
send the graph and the links' coloring into themselves. See Appendix 
C for an example, and reference \cite{gi,ashtekar3} for a rigorous 
construction.   It is useful to view an s-knot state as a 
group integral of a spin network state:
\begin{equation}
	\langle s| = {\cal N} \int_{Dif\!f_0}[df]\ \langle f\cdot S|,
\label{sknotint}
\end{equation}
where $S\in s$. 

\subsection{The hamiltonian constraint}

The Hamiltonian constraint that we consider is the density-weight 1 
hamiltonian density, smeared with a density-weight 0 Lapse function $N$.  
The Lorentzian hamiltonian constraint $C_L[N]$ can be written as the sum of 
two terms: $C_L[N]=C[N]+V[N]$ \cite{barbero}, where $C_L[N]$ is the 
Euclidean Hamiltonian constraint.  For simplicity we deal here only with 
the first term.  Thus, we are dealing below with Euclidean quantum gravity 
only.  We expect the methods developed here to be extendible to the 
$V[N]$ term as well, and therefore to Lorentzian GR, using the techniques 
developed by Thiemann \cite{Thiemann}.

The definition of $C[N]$ is plagued by ordering ambiguities
\cite{hamiltonian,Thiemann,outline}.  Some of these are fixed by 3d diff
invariance \cite{hamiltonian}. In section V we discuss how 4d diff
invariance might fix others. Here, we recall Thiemann's version of the 
Hamiltonian constraint, which is the starting point of our construction.
First, the non-symmetric operator $C_{ns}[N]$ is defined as
 \begin{equation}
  	C_{ns}[N]\ |S\rangle = 
	2 \sum_{i\in n(S)} 
	N(x_i)
	\sum_{(JK)\in e(i)}\  
	\sum_{\epsilon=\pm 1,\epsilon^\prime= \pm 1} 
	A_{iJK\epsilon\epsilon'}(S)\  
	 D_{iJK\epsilon\epsilon'}\ |S\rangle.
\label{c}
\end{equation}
Here $i$ labels the nodes of $S$ (which form the set $n(S)$); 
$x_i$ are the coordinates of the node $i$; 
$(JK)$ labels the couples of distinct links adjacent to 
the node $i$ (these form the set $e(i)$); the operator
$D_{iJK\epsilon\epsilon\prime}$ was introduced in \cite{outline}, 
it acts on the spin network by 
creating two new trivalent nodes $i'$ and $i''$ on the the two 
links $J$ and  $K$ respectively, connected by a link with color 
$1$, and adds  $\epsilon$ (resp. $\epsilon'$) to the color of the 
link connecting $i$ and $i'$ (resp. $i$ and $i''$).  This is 
illustrated in Figure 1. 
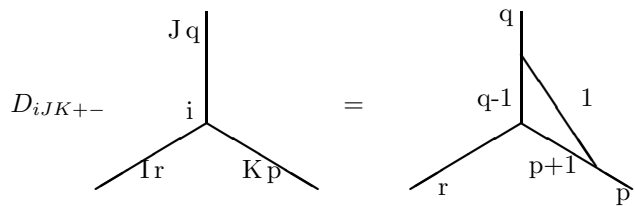
\begin{figure}
\setlength{\unitlength}{0.7pt}
\centerline{\mbox{
\begin{picture}(344,114)(105,606)
\thicklines
\put(201,648){\line(-5,-3){ 60}}
\put(201,648){\line( 5,-3){ 60}}
\put(201,648){\line( 0, 1){ 60}}
\put(165,618){\makebox(0,0)[lb]{\raisebox{0pt}[0pt][0pt]{I\,r}}}
\put(180,694){\makebox(0,0)[lb]{\raisebox{0pt}[0pt][0pt]{J\,q}}}
\put(220,618){\makebox(0,0)[lb]{\raisebox{0pt}[0pt][0pt]{K\,p}}}
\put(190,650){\makebox(0,0)[lb]{\raisebox{0pt}[0pt][0pt]{i}}}
\put(95,654){\makebox(0,0)[lb]{\raisebox{0pt}[0pt][0pt]
{$D_{iJK+-}$}}}
\put(371,685){\line( 2,-3){ 40.462}}
\put(371,648){\line(-5,-3){ 60}}
\put(371,648){\line( 5,-3){ 60}}
\put(371,648){\line( 0, 1){ 60}}
\put(275,654){\makebox(0,0)[lb]{\raisebox{0pt}[0pt][0pt]{=}}}
\put(326,609){\makebox(0,0)[lb]{\raisebox{0pt}[0pt][0pt]{r}}}
\put(422,606){\makebox(0,0)[lb]{\raisebox{0pt}[0pt][0pt]{p}}}
\put(359,703){\makebox(0,0)[lb]{\raisebox{0pt}[0pt][0pt]{q}}}
\put(403,657){\makebox(0,0)[lb]{\raisebox{0pt}[0pt][0pt]{1}}}
\put(347,657){\makebox(0,0)[lb]{\raisebox{0pt}[0pt][0pt]{q-1}}}
\put(375,621){\makebox(0,0)[lb]{\raisebox{0pt}[0pt][0pt]{p+1}}}
\end{picture}
}}
\caption{Action of $D_{iJK\epsilon\epsilon'}$. $r, q$ and $p$ are the 
colors of the links $I, J$ and $K$.} \end{figure} The precise location of 
the nodes and the link added is an arbitrary regularization choice.  
The coefficients $A_{iIJ\epsilon\epsilon^\prime}(S)$ 
of Thiemann's operator are well defined and can be computed  
explicitly \cite{rrr}.  They are functions of the colors of the links 
adjacent to the node $i$; they are finite and can be expressed as 
products of linear  combinations of n-$j$ symbols of $SU(2)$.  

It is important to notice that Thiemann's operator was derived using the 
infinite-dimensional differential-geometry techniques introduced in
\cite{ashtekar}.  These differential techniques were introduced as a 
mathematical sistematization of the ideas on loop quantization introduced 
in \cite{loops}.   They have shed much light on loop quantum 
gravity, have provided a rigorous 
mathematical foundation of the theory, and have led, among other 
results, to Thiemann's operator.  However, the operator itself is 
a well-defined {\it algebraic\/} operator on the spin network basis, and the 
computation of its matrix elements is easier using algebra than using 
infinite-dimensional differential-geometry \cite{dpr}.  The equivalence 
between the purely algebraic formalism (``loop representation'' or ``spin 
network representation'') and the differential formalism (``connection 
representation'') is shown in detail in \cite{rob}
(see also \cite{inverse,volume}). The situation is 
analogous to the two well-known ways of computing the spectrum of the 
harmonic oscillator: one can use Dirac's {\it algebraic\/} technique, in 
the $|n\rangle$ basis; or, alternatively, one can solve the {\it 
differential\/} Schr\"odinger equation, in the coordinate basis.  Each 
basis has its own advantages, but the two formalisms are equivalent, 
and there is no sense in which one representation is more ``rigorous" than 
the other.  In particular, spectra of area and volume in quantum 
gravity can be computed either using algebraic techniques (this is the way 
they were {\it first\/} computed and their discreteness was discovered in 
\cite{discr}) or using differential techniques (see \cite{volume}, and 
references therein).  The resulting spectra are, of course, equal \cite{fl}. 

We shall use Thiemann's {\it symmetric\/} hamiltonian constraint 
operator $C[N]$, defined by symmetrizing $C_{ns}[N]$ 
\begin{equation} 
\langle S' |C[N]| S \rangle = 
{1\over 2}
\left(\ \langle S' | C_{ns}[N]| S \rangle\ +\ \langle S |C_{ns}| S' 
\rangle\ \right).
\label{h2}
\end{equation}
Explicitly, we have 
\begin{eqnarray} 
	\langle S' |C[N]| S \rangle &=& 
	\sum_{i\in n(S)} N(x_i)
	\sum_{(JK)\epsilon\epsilon\prime} 
	A_{iJK\epsilon\epsilon\prime}(S)\  
	\langle S' |D_{iJK\epsilon\epsilon\prime}|S\rangle
\n \\ 	&& 
+ \sum_{i'\in n(S')} N(x_{i'})	
	\sum_{(JK)\epsilon\epsilon\prime} 
	A_{i\prime JK\epsilon,\epsilon'}(S)
	\langle S' |D^\dagger_{i\prime JK\epsilon\epsilon'}|S\rangle
\label{h3}
\end{eqnarray}
where  
\begin{equation}
	\langle S' |D^\dagger_{iJK\epsilon\epsilon'}|S\rangle
	=\langle S |D_{iJK\epsilon\epsilon'}|S'\rangle.
\end{equation}

An observation important for what follows is that the hamiltonian 
constraint is ``local'', in the following 
sense. Given a spin network $S$, we may cut it in two parts, by 
cutting  $n$ links, obtaining two spin networks with open ends, which
we denote $\tilde S$ and $\hat S$.  Imagine we have two spin networks 
$S_i$ and $S_f$ that can be cut as $\tilde S_i$ and $\hat S_i$, and, 
respectively, $\tilde S_f$ and $\hat S_f$. Imagine that $\hat 
S_i=\hat S_f$, namely the two spin networks differ only in their
``tilde'' component.  Then the matrix elements  $\langle 
S_f|C[N]|S_i\rangle$ do not depend on the ``hat'' components 
$\hat S_i$ and $\hat S_f$, so we can write
\begin{equation}
\langle S_f|C[N]|S_i\rangle = \langle \tilde S_f|C[N]|\tilde 
S_i\rangle. 
\label{tilde}
\end{equation}
This decomposition will play an important role below.

We simplify notation by introducing a single discrete index $\alpha, 
\beta \dots $ to replace the discrete set of indices 
$(i,JK,\epsilon,\epsilon^\prime)$.  For every spin network $S$, 
$\alpha$ ranges over a finite set $[S]$ of values, with
 \begin{equation}
4\prod_{i\in n(S)} {v_i(v_i-1)\over2}
\end{equation}
values, where $v_i$ is the valence of the node $i$. We also 
indicate by $x_\alpha$ the coordinates of the node with index 
$\alpha$.  Using  this, we have
\begin{equation}
        \langle S' |C[N]| S \rangle =
        \sum_{\alpha\in [S]} N(x_\alpha)\ 
        A_\alpha(S)\
        \langle S' |D_\alpha|S\rangle
+ \sum_{\beta\in [S']} N(x_\beta)\ 
       A_\beta(S)\ 
        \langle S' |D^\dagger_\beta|S\rangle. 
\label{h4}
\end{equation}

The hamiltonian constraint transforms covariantly under the 
diffeomorphisms generated by the diffeo constraint
\begin{equation}
\left[C[\vec N],C[N]\right]=C[{\cal L}_{\vec N}N], 
\end{equation}
where ${\cal L}_{\vec N}$ is the Lie derivative along $\vec N$.  
Under a finite diffeomorphism $f$, we have
\begin{equation}
	D[f]\ C[N]\ D^{-1}[f]=C[N_{f}],
\label{14}
\end{equation}
where $N_f$ is the transformed Lapse:
\begin{equation}
	N_{f}(x) \equiv N(f(x)).
\end{equation}
The transformation properties of $C[N]$ under 4d diffeomorphisms are less
clear.  In the canonical formalism these are controlled by the commutator
of $C[N]$ with itself, which, however, is not fully under control, due to
the interplay between regularization and 3d diff invariance
\cite{bernie2}. (Notice that in \cite{Thiemann} it is shown that the
commutator $[C_{ns}[N],C_{ns}[M]]$ vanishes on diff invariant states; the
commutator $[C[N],C[M]]$ is more tricky.) In Section IV we suggest a way
for addressing these difficulties. 

\section{Proper time evolution operator $U(T)$}

Consider, as an illustrative example, the Schr\"odinger equation for a 
single particle in a potential.  If $H$ is the Hamiltonian operator, 
the equation is formally solved by the evolution operator
\begin{equation}
	U(t) = U(t,0) = e^{-i\int_0^t dt' H(t')}
\end{equation}
where exponentiation, here and below, is time ordered.  The matrix elements 
of this operator between position eigenstates define the propagator
\begin{equation}
	P(\vec x,t;\vec x',t') = \langle \vec x|U(t,t')|\vec x\rangle,
\end{equation}
from which the solution of the Schr\"odinger equation with initial data 
$\psi(\vec x', t')$ at $t'$ can be obtained by simple integration
\begin{equation}
	\psi(\vec x,t) = \int dx' P(\vec x,t;\vec x't') \psi(\vec x',t'). 
\end{equation}
Under suitable conditions, the propagator can be computed by means 
of a perturbation expansion in the potential, and the expansion 
has a nice graphical representation.

For a (free) relativistic particle, we have the option between using 
the above formalism with the relativistic Hamiltonian 
($H=\sqrt{\vec p^2+m^2}$), or using a manifestly covariant formalism. 
This was originally done by Feynman by changing the description of the 
dynamics: instead of representing motion by means of the evolution of the 
{\it three} variables $\vec x$ in $t$, we consider a (``fictitious'') 
evolution of the {\it four} variables $x=(\vec x, t)$ in the proper time 
$T$.  This evolution is generated by the operator $H=p^2-m^2=(p^0)^2-\vec 
p^2-m^2$.  The corresponding proper time evolution operator and proper 
time propagator are
\begin{equation}
	U(T) = e^{-i\int_0^T dt' H(t')}
\label{Ut}
\end{equation}
and
\begin{equation}
   P(\vec x,t;\vec x',t';\, T) = \langle \vec x, t |U(T)|\vec x', 
t'\rangle. 
\end{equation}
The relation between this proper time propagator and the physical 
propagator (which is the quantity we compare experiments with) is 
given by 
\begin{equation}
   P(\vec x,t;\vec x',t') = \int_0^\infty dT\  P(\vec x,t;\vec x',t';\,T).
\end{equation}
Or
\begin{equation}
  P(\vec x,t;\vec x',t') = \langle \vec x, t |U|\vec x', t'\rangle, 
\end{equation}
where
\begin{equation}
	 U = \int_0^\infty dT \ U(T).
\label{proptimeint}
\end{equation}
This can be verified by means of a simple calculation.   $U$ is the 
projector on the physical state space, which codes the theory's dynamics.

Alternatively, one can consider evolution in a fully arbitrary 
parameter $t$.  Such evolution is generated by $H(t)=N(t)H$, where $N(t)$ 
is an arbitrary Lapse function.  The corresponding evolution operator is
\begin{equation}
	U_N= e^{-i\int_0^1 dt N(t) H},
\end{equation}
which is related to the physical $U$ by the functional integral
\begin{equation}
	 U = \int [dN]\ U_N.
\label{fi}
\end{equation}
This functional integration can be split into two parts by defining 
the proper time $T$ in terms of the Lapse as
\begin{equation}
	T = \int_0^1 N(t)\ dt.
\label{T}
\end{equation}
Using this, we can first integrate $U_N$ over all Lapses $N$ having 
the same $T$. 
\begin{equation}
\tilde U(T)=\int_T dN\ U_N
\label{fix}
\end{equation}
where the subscript $T$ indicates that the functional integral must be 
performed over all $N$'s satisfying \ref{T}.
Then we  integrate over $T$ to get the physical quantity $U$.  Thus we have
\begin{equation}
U_N \ \ \ \longmapsto \ \ \ \tilde U(T) \ \ \  \longmapsto  \ \ \ U. 
\end{equation}
  
An important observation is 
that in order to compute the functional integral \ref{fix} we can simply 
{\em gauge fix\/} $N$, requiring, for instance $dN(t)/dt=0$. In this 
gauge, the  integral becomes trivial, and we have $\tilde U(T)= U(T)$, 
which is given in \ref{Ut}.  In fact, the functional integration 
\ref{fi} over $N$ is largely trivial, since $U_N$ depends on $N$ only 
via $T$. 

We are now going to follow the same path in general relativity.  In 
particular, we will concentrate here on the definition and the computation 
of the proper time evolution operator $U(T)$ and the corresponding proper 
time propagator (its matrix elements) for quantum general relativity.

\subsection{Definition and meaning of the proper time propagator in general 
relativity}

In the canonical theory, the (``unphysical'' or coordinate-) 
evolution of the gravitational field is generated by the hamiltonian
\begin{equation}
H_{N,\vec N}(t)=
\int_\Sigma d^3x\ [N(t,x) C(x) + N^a(t,x) C_a(x)]
\equiv C[N(t)]+C[\vec N(t)] \label{hamiltonian} 
 \end{equation} 
(units are fixed here by $\hbar=c=16\pi G_{Newton}=1$, and we take $\Sigma$ 
compact).  The quantum evolution operator that evolves from an initial 
hyper-surface $\Sigma_{i}$ at $t=0$ to a final hyper-surface $\Sigma_{f}$ 
at $t=1$ is
\begin{equation}
  U_{N,\vec N} = e^{-i \int_0^1 dt\ H_{N\vec N}(t)}.
\label{zu}
\end{equation}
We define the proper time evolution operator for quantum gravity as
\begin{equation}
U(T) = \int_{T,*} [dN,\, d\vec N]\ \ U_{N, \vec N}\ \ ,
\label{uvero}
\end{equation}
where the subscript $\{T,*\}$ means that the integral is over all Shifts 
and Lapses that satisfy
\begin{eqnarray}
N(x, t) & = & N(t)  
\label{star}   \\
\int_0^1 dt\ N(x,t) & = & T. 
\label{ttt}
\end{eqnarray}
Notice that $T$ is the proper time separation between $\Sigma_{i}$ and 
$\Sigma_{f}$, defined as the reading on $\Sigma_{f}$ of the free falling 
test clocks that started off at rest on $\Sigma_{i}$.  This is because 
if the Lapse is constant the geodesics that define the proper time 
foliation remains normal to 
the ADM hypersurfaces.  We denote the matrix elements of the operator 
(\ref{uvero})
\begin{equation}
        P(s_f, s_i;\, T) = \langle s_f |\, U(T)\, | s_i \rangle 
\label{ptp}
\end{equation}
as the proper time propagator \cite{claudio}.  In this paper, we 
focus on this quantity.  We compute it as a power expansion in $T$ in 
the next subsection, and show that it admits a sum over surfaces 
representation in section IV.
\footnote{Generally, slicing by equal proper time hypersurfaces 
develops singularities, because the geodesics that define the proper time
intersect. This causes the canonical evolution to break down: a coordinate
system based on the slicing develops coordinate singularities where the
ADM momentum density diverges. We ignore these difficulties here, but two
comments are in order.  First, the explicit expression for $U(T)$ that we
obtain is simple and well defined order by order for any $T$. We think
that potential singularities in $U(T)$ should be looked for directly in
the quantum formalism. Second, Lewandowski \cite{Jurek} has pointed out
that the classical evolution of the Ashtekar's variables is better behaved
than the ADM variables, because Ashtekar's variables can be represented as
differential forms, whose components are well behaved at coordinate
singularities of the type being considered.}

The construction generalizes to a multifingered proper time.  
In this case, let $\Sigma_{f}$ is given by $t=t(x)$.  The coordinate 
time evolution operator from $\Sigma_{i}$ to $\Sigma_{f}$ at fixed 
Lapse and Shift is
\begin{equation}
  U_{N,\vec N} = e^{-i \int_\Sigma d^3x \int_0^{t(x)} dt\ [N(t,x) C(x) + 
  N^a(t,x) C_a(x)]}.
\end{equation}
The multifingered proper time evolution operator is
\begin{equation}
U[T] = \int_{[T]} [dN, \,d\vec N]\ \ U_{N, \vec N}\ \ ,
\label{uvero2}
\end{equation}
where the subscript $[T]$ means that the integral is over all Shifts 
and Lapses that satisfy (\ref{star}) and
\begin{equation}
\int_0^{t(g^{-1}_t(x))}  dt\ N(g^{-1}_t(x),t) = T(x)
\label{tttt}
\end{equation}
where $g_t$ is the finite, time-dependent, transformations of spatial 
coordinates generated by integrating the shift:
\begin{equation}
 g_0(x)=x, \ \ \  {d g_t(x)\over dt} = \vec N(x, t),  \ \ \ g \equiv g_1.
\label{g}
\end{equation}
$g_t$ and $g$ are functionals of the Shift.  As before, $T(x)$ gives 
the proper time separation between the two hyper-surfaces, defined as 
the proper distance of $\Sigma_{f}$ from $\Sigma_i$ along a geodesic 
starting at rest on $\Sigma_i$ on $x$ Indeed, if the Lapse is spatially 
constant, the geodesic will be $g^{-1}_{t}(x)$ (because in 
the coordinates $(\vec y, t) = (g_{t}(x), t)$ the Shift vanishes, the 
Lapse is still constant and therefore $\vec y = $ constant is a 
geodesic normal to all ADM slices).  And therefore the geodesic that 
starts off in $x$ reaches $\Sigma_{f}$ at the time $t$ determined by 
$t=t(g_{t}^{-1}(x))$.  Notice that the two notations $U[T]$ and $U(T)$
indicate different objects.  $U[T]$ is the {\em multifingered\/} proper
time propagator, a functional of the function $T(x)$; while $U(T)$ is the
proper time evolution operator, which is the value of $U[T]$ for $T(x)=
constant = T$. 

In the rest of this subsection, we discuss the physical meaning of the
quantity we have defined and its role in the theory. 

First of all, the proper time propagator is the first step toward the 
computation of the physical evolution operator $U$, in the same sense as
the Feynman-Nambu proper time propagator.  The  operator 
$U$ is given by functionally integrating $U_{N,\vec N}$ over {\it all\/} 
lapses and shifts:
\begin{equation}
	 U = \int [dN] \int [d\vec N] \ U_{N,\vec N}.
\label{fi2}
\end{equation}
This functional integrations corresponds to the implementation of the 
canonical constraints.  As for the relativistic particle considered in 
the previous section, we can split the computation of $U$ from 
$U_{N,\vec N}$ into two steps
\begin{equation}
U_{N, \vec N} \ \ \ \longmapsto \ \ \ \tilde U[T] \ \ \  \longmapsto  \ \ 
\ U, \end{equation} 
by first computing the propagator $\tilde U[T]$ at fixed $T(x)$ and then
integrating over $T(x)$. As for the particle, we can partially fix the
gauge in which we compute $U[T]$. In particular, we can choose to
integrate over spatially constant lapses only. Therefore we have $\tilde
U[T]= U[T]$, which is given in \ref{uvero2}.  And, as for the particle,
we can write
\begin{equation}
 U = \int [dT] \ \ U[T]
\end{equation}
Thus, $U$ is just the integral over proper time of the multifingered 
proper time evolution operator (\ref{uvero2}).

We add a general argument that better illustrates why we can fix a 
gauge for computing $\tilde U[T]$.  This argument is formal, but it 
is interesting because it illuminates the relation between what we are 
doing and the sum over geometries considered by Hawkings \cite{hh}.
In the metric formulation of canonical GR, $U$ can be written as a 
sum over four-metrics bounded by given initial and final 
three-geometries. Each such four-metric determines a proper time 
separation $T(x)$ between the initial and final hypersurfaces. 
Therefore, the integral can be split in two parts, first the integral 
$\tilde U(T)$, restricted to 4-metrics with total multifingered elapsed time 
$T(x)$, then the integration over $T(x)$. 
In computing $U(T)$, we can change integration variables from the 
four-metric to the ADM variables, namely three-metric, Lapse 
and Shift. The integral contains a high 
redundancy, corresponding to diffeomorphism gauge-invariance (as the 
corresponding integral for the particle in the previous section did). 
We can fix part of this redundancy with a condition 
on the Lapse.  If we pick an arbitrary Lapse, the condition that the 
proper time separation of the initial and final slices is $T(x)$  becomes a 
condition on the three-metrics over which we are integrating.  However, 
we can choose a spatially constant Lapse, satisfying \ref{star} and 
\ref{tttt}. We can always do that, because we may always slice a four 
geometry with equal proper time hypersurfaces, with the result that the 
corresponding Lapse and Shift satisfy the (\ref{star}) and 
(\ref{tttt}).  Conversely, any history of 3-metrics,  together with a 
Lapse and Shift satisfying (\ref{star}) and (\ref{tttt}) defines a 
4-geometry with elapsed proper time $T(x)$.  In this way, we implement 
the $T(x)$ condition, without interfering with the integration over 
three-metrics. Thus, $\tilde U[T]$ can be computed by 
fixing a Lapse satisfying (\ref{star}) and (\ref{tttt}). 

The operator $U$ is a key quantity for the theory.  Computing it 
virtually amounts to solving the quantum constraints, including the 
hamiltonian constraint, which codes all the dynamics of the theory.  
There are various of ways of looking at $U$.  First of all, it is the 
projector on the the physical state space of the theory.  Second, the 
scalar product $\langle s | U | s' \rangle$ defines the {\it 
physical\/} scalar product of the theory.  Therefore
\begin{equation}
\langle s, \ s' \rangle_{{\rm physical}} = \int [dT] \ 
\langle s |\, U[T]\,|\ 
s' \rangle.
\end{equation}
Finally, we can view matrix elements of $U$ as observable 
transition amplitudes between quantum states.  The details of the 
interpretation of $U$ will be discussed elsewhere, but in all these 
instances, the role of $U$ is just analogous to its counterpart for the 
single particle.  (In this paper we do not attempt to compute $U$.)

Does the proper time propagator $P(s_f, s_i;\, T)$, have a direct  
physical interpretation?  A simple answer is that $P(s_f, s_i;\, T)$ 
codes the dynamics of the theory, but it has no direct physical 
meaning: only after integration over proper time we obtain a quantity 
that we can, in principle, campare with experiments.

This said, we {\it can\/} nevertheless assign a plausible physical 
interpretation to the proper time propagator $P(s_f, s_i;\, T)$, with a 
certain caution.  This would be particularly useful for helping 
intuition.  Let us return to the relativistic particle.  In that case, 
Feynman considered the ``fictitious'' evolution of $x$ and $t$ in the 
proper time $T$.  Classically, this is not incorrect (because the 
equations of motion of the 4+1 dimensional theory give the correct 
physical 3+1 motion) provided that one remembers that the degrees of 
freedom are 3 and not 4.  Quantum mechanically, in the fictitious 
theory we are quantizing one variable too much.  Taken literally, the 
particle proper time propagator describes the 3 degrees of freedom of 
the particle position, plus the degree of freedom of an extra quantum 
variable sitting on the particle, growing with proper time, and not 
affecting the particle's motion\footnote{More precisely, the extra 
variable is not the evolution of this proper-time clock-variable $T$ 
in a Lorentz time $x^{0}$, but rather the evolution of $x^{0}$ in $T$.}.  
With such a little clock on the particle (say the particle is an oscillating 
molecule), we could make experiments we could compare the proper time 
propagator with.

In general relativity, such a ``fictitious'' evolution with extra 
degrees of freedom is provided by the so-called ``local 
interpretation'' of the theory (see \cite{obs} for a detailed 
discussion).  In this interpretation, the coordinates are interpreted as  
labels of reference-system physical objects (RS-objects).  It follows 
that local quantities are physical observables, and that the lack of 
determinism of the Einstein equations can be interpreted as a consequence 
of the fact that the dynamical equations of the RS-objects are 
neglected.  Under this interpretation, GR is approximate (because we 
disregard the RS-objects energy momentum) and incomplete (because we 
disregard the RS-objects dynamical equations).  The incompleteness 
leads to the apparent physical indeterminism.  If we adopt this view, 
then we can say that $s, s'$ and $T$ are observable, because 
$\Sigma_{i}$ can be physically specified by the RS-objects, and we can 
use RS-clocks to find out where is $\Sigma_{f}$.  In doing so, we take 
approximations that might be ungranted (on the quantum behavior of the 
RS-objects).  Concretely, one may consider a definite model, for 
instance the ``dust'' model introduced in \cite{obs} and studied in 
\cite{kk}.  In such a model, a $\Sigma_{i}$ to $\Sigma_{f}$ propagator 
(where $\Sigma_{i}$ is defined dust variables) is an observable 
quantity.  In a suitable limit in which the dust physical effects are 
disregarded, such propagator might be approximated by the pure gravity 
proper time propagator $P(s_f,s_i;\, T)$.

With all these caveats, one can intuitively think of $P(s_f,s_i;\, T)$ as 
the quantum  amplitude that the quantum gravitational field be in the 
state $s_f$, $T$ seconds after being in the state $s_i$.

\subsection{Expansion of the proper time propagator}

We begin with an observation.  Let us write $U_{N,\vec N}$ as a 
limit of products of small time propagators.  Writing $\epsilon=1/K$ 
and $t_k=k\epsilon$, for integers $K$ and $k=1 ...  K$, we have
\begin{eqnarray}
U_{N,\vec N}&=&\lim_{K\rightarrow\infty} e^{-i\epsilon H_{N,\vec N}(t_K)} 
\dots \ 
e^{-i\epsilon H_{N,\vec N}(t_2)}\  
e^{-i\epsilon H_{N,\vec N}(t_1)} 
\n \\
&=& \lim_{K\rightarrow\infty} 
e^{-i\epsilon C[N(t_K)]}\, e^{-i\epsilon C[\vec N(t_K)]} \dots \  
e^{-i\epsilon C[N(t_2)]}\, e^{-i\epsilon C[\vec N(t_2)]}  \ \ \ 
e^{-i\epsilon C[N(t_1)]}\, e^{-i\epsilon C[\vec N(t_1)]}
\n\\
&=& \lim_{K\rightarrow\infty}
e^{-i\epsilon C[N(t_K)]} D[f_K] \dots \ 
e^{-i\epsilon C[N_{t_2}]}\, D[f_2]\ 
e^{-i\epsilon C[N_{t_1}]}\, D[f_1]
\n\\
&=& \lim_{K\rightarrow\infty}
e^{-i\epsilon C[N(t_K)]} D[f_K] \dots 
\left(
(D[f_2]D[f_1])^{-1}
e^{-i\epsilon C[N(t_2)]}  D[f_2]D[f_1] \right)
\left(D^{-1}[f_1] e^{-i\epsilon C[N(t_1)]} D[f_1]\right) 
\n \\ 
&=& \lim_{K\rightarrow\infty}
D[g] \left(D^{-1}[g_K] e^{-i\epsilon C[N(t_K)]} 
D[g_K]\right) 
\dots\  
\left(D^{-1}[g_2] e^{-i\epsilon C[N(t_1)]} D[g_2]\right) 
\left(D^{-1}[g_1] e^{-i\epsilon C[N(t_1)]} D[g_1]\right) 
\n\\
&=& \lim_{K\rightarrow\infty}
D[g]\ \  e^{-i\epsilon C[N_{\vec N}(t_K)]} \dots\ 
e^{-i\epsilon C[N_{\vec N}(t_2)]} 
e^{-i\epsilon C[N_{\vec N}(t_1)]} 
\n \\ 
&=& D[g]\ \  U_{N_{\vec N},0}.
\label{observation}
\end{eqnarray}
Here $N_{\vec N}$ is defined by $N(x, t) = N(g_{t}(x), t)$, namely it 
is the lapse in the coordinates obtained by integrating the shift.  
$f_t$ is the ``small'' diffeomorphism generated by the Shift between 
the slices $t-1$ and $t$; while $g_t$ is the finite diffeomorphism 
generated by the Shift between the slices $t=0$ and $t$.  The first 
equality in (\ref{observation}) is just one of the definitions of the 
time ordered exponential.  The second is based on the fact that for 
sufficiently small time interval $1/K$ (sufficiently high $K$) one can 
disregard the commutator term in disentangling the exponent (this term 
is quadratic in $1/K$).  The third equality is simply a rewriting of 
the exponent of an infinitesimal diffeomorphism as a finite (but 
``small'') diffeomorphism.  The fourth equality is simply the 
insertion of terms like $(D[f_1]D^{-1}[f_1] )$ in suitable places.  
The fifth equality is the replacement of sequences of spatial 
diffeomorphisms $(D[f_n] ...  D[f_2] D[f_1])$ by their product, which 
is $(D[g_n])$.  The penultimate equality is the key one; it follows 
directly from equation (\ref{14}), namely from the transformation 
properties of the hamiltonian constraint under spatial 
diffeomorphisms.  The last equality follows again from the definition 
of ordered exponential.  In words, we have shown that the temporal 
evolution generated by the Lapse and the evolution generated by the 
Shift can be disentangled.

While the manipulations above are formal (they are made inside a 
limit), the result itself is geometrically obvious: we can always 
rearrange the coordinates so that the Shift is zero, and compensate with a 
finite change of space coordinates at the end.  If we do so, the Lapse $N$ 
must be replaced by the Lapse in the new coordinates, which is $N_{\vec N}$.

If the lapse is constant in space, $N_{\vec N}=N$.  Then $U_{N,0}$ can be 
expanded as
\begin{equation}
U_{N0} = 1 + (-i) \int_0^\tau\! dt\ C[N(t)]
+ (-i)^2 \int_0^\tau\! dt \int_t^\tau\! dt'\ C[N(t')]\ C[N(t)]
+\ \dots \ .
\end{equation}
Its matrix elements between two spin network states can be 
expanded as
\begin{equation}
\langle S_f|U_{N,0}(T)|S_i\rangle = 
\langle S_f|S_i\rangle  + (-i) \int_0^\tau\!dt 
\langle S_f|C[N(t)]|S_i\rangle 
+ (-i)^2 \int_0^\tau\! dt \int_t^\tau\! dt'\  
\langle S_f|C[N(t')]|S_1\rangle\ \langle S_1|C[N(t)]|S_i\rangle 
+ \dots
\end{equation}
 where we have inserted a complete set of intermediate states
$|S_1\rangle\ \langle S_1|$ (over which summation is understood). 
Using the explicit form (\ref{h4}) of the hamiltonian constraint 
operator, we have
 \begin{eqnarray}
\langle S_f|U_{N,0}|S_i\rangle &=&
\langle S_f|S_i\rangle \n\\ 
&& + (-i) \int_0^\tau dt\
        \left(\sum_{\alpha\in [S_i]} 
	N(t, x_\alpha)
        A_\alpha(S_i)\ 
        \langle S_f | D_\alpha|S_i\rangle 
       +     \sum_{\beta\in[S_f]} 
	N(t,x_\beta)
        A_\beta(S_f)\ 
        \langle S_f | 
D^\dagger_\beta|S_i\rangle \right)
\n \\ && + (-i)^2 \int_0^\tau\! dt \int_t^\tau\! dt'
   \sum_{\alpha\in [S_i]}   \sum_{\alpha'\in [S_1]}
N(t,x_\alpha) N(t',x_{\alpha^\prime})\ 
A_\alpha(S_i)  
A_{\alpha^\prime}(S_1) \ 
 \langle S_f | D_\alpha|S_1\rangle 
\langle S_1 | D_{\alpha^\prime}|S_i\rangle
\n \\   & & + \ \ \ \dots
\end{eqnarray}
(the second order term has three more summands, corresponding to 
the $DD^\dagger, D^\dagger D, D^\dagger D^\dagger$ terms). The 
first point to be noticed in this expression is that the sum over the 
intermediate state $S_1$ is finite.  This is because both $D$ and 
$D^\dagger$ yield a finite number of terms only, when acting on a spin 
network state.\footnote{
Lewandowski \cite{Jerzy} has noticed that this finiteness might fail 
because of the moduli parameters of high valent intersections which were 
studied in \cite{norbert}. The role of these parameters in the theory,
however, is unclear. Finiteness of the proper time expansion
may indicate that the correct version of the
theory is the one in which the moduli parameters are removed, as 
suggested by many, and recently detailed in \cite{Zapata}.}
 Thus, the above expression is finite order by order.  
Next, the integrations can be performed explicitly, using (\ref{ttt}).  
We obtain \begin{eqnarray}
\langle S_f|U|S_i\rangle &=&
\langle S_f|S_i\rangle \n \\ && 
+ (-iT) \left(
        \sum_{\alpha\in[S_i]} A_\alpha(S_i)\
        \langle S_f |D_\alpha|S_i\rangle +
        \sum_{\alpha\in[S_f]} 
         A_\alpha(S_f)\ 
        \langle S_f | 
D^\dagger_\alpha|S_i\rangle \right)
\n \\ && + {(-iT)^2\over 2!} 
   \sum_{\alpha\in[S_i]}   \sum_{\alpha\prime\in[S_1]}
  A_\alpha(S_i)\  A_{\alpha\prime}(S_1)\ 
 \langle S_f |\hat D_{\alpha\prime}|S_1\rangle \ 
\langle S_1 |\hat D_\alpha|S_i\rangle
\n \\ 
 & &  + \ \ \  \dots \ .
\label{expansion}
\end{eqnarray}
The structure of the expansion is now rather clear. At each order 
$n$, we have the $D$ operator acting $n$ times, $n$ factors $A$, and a {\it 
finite} number of terms, coming from summing over nodes, links 
and $\epsilon=\pm1$.  

Our next step is to integrate over Shift and Lapse (satisfying
(\ref{tttt})).  The integration over lapse is trivial, as its dependence 
has dropped out the integral.  This confirms the independence from the 
lapse that was mentioned in the previous section. 
 The integration over the Shift amounts to imposing the 
Diff constraint.  Indeed, it turns out to be equivalent to an 
integration over the diffeomorphism group, as in the group integration 
technique for solving the Diff constraint.  Using (\ref{observation}), 
we have
\begin{equation}
  U(T) = {\cal N}^\prime\int [dN]\int [d\vec N]\ D[g[\vec N]]\ 
U_{N_{\vec N}, 0}(T), 
\end{equation}
where we have explicitly indicated the dependence of $g$ on $\vec 
N$ for clarity. We change integration variable $N\rightarrow N_{\vec N}$ 
(the Jacobian must be one, since this amounts to a change of coordinates  
)), and obtain
\begin{equation}
  U(T) = {\cal N}^\prime\left(\int [dN]\ U_{N,0}(T)\right)\left(\int 
[d\vec N]\  D[g[\vec N]]\right) .  
\end{equation}
The $\vec N$ integration can be traded for an integration 
over $Dif\!f_0$ changing variables from $\vec N$ to $g[\vec N]$, so we 
obtain
 \begin{equation}
  U(T) = {\cal N} \int_{Dif\!f_0} [dg]\ D[g]\ U_{N,0}(T)
 \end{equation}
for an arbitrary (irrelevant) choice of $N$ satisfying (\ref{ttt}), say 
$N=T$. The matrix elements of this operator are given by 
\begin{equation}
\langle S_f |U(T)| S_i \rangle 
= {\cal N}\int_{Dif\!f_0}[dg]\ \ \langle g\cdot S_f |U_{N,0}(T)| S_i 
\rangle. 
\end{equation}

The operator $U(T)$ is now well defined in ${\cal H}_{diff}$! Indeed, it is 
immediate to see that it is diff-invariant.  For every two s-knots 
$s_i$ and $s_f$ in ${\cal H}_{diff}$, we can arbitrarily pick $S_i$ and 
$S_f$ 
such that $S_i\in s_i$ and $S_f\in s_f$, and we have the key result that
\begin{equation}
\langle s_f |U(T)| s_i\rangle \equiv \langle S_f |U(T)| S_i\rangle 
\label{sp}
\end{equation}
is well defined (independent from the $S_i$ and $S_f$ chosen). 

Furthermore, the operator $D$, depends on an arbitrary 
regularization --the location of the added link--, but a moment of 
reflection shows that the dependence on the regularization drops out in the 
step from $U_{N,\vec N}(T)$ to $U(T)$, by integrating the Shift.  The 
reason is that different regularizations are related to each other by a 
finite diffeomorphism: the states $D_\alpha|S\rangle$ and 
$D^\prime_\alpha|S\rangle$, where $D$ and $D^\prime$ indicate two different 
regularizations of $D$ are in the same s-knot: their difference becomes 
irrelevant in the scalar product (\ref{sp}).  This result is due to the 
fact that all the factors in the expansion are individually well 
defined at the diffeomorphism invariant level.  More precisely we have that
 \begin{equation}
\langle s_f|\sum_{\alpha\in[s_i]} A_\alpha(s_i)D_\alpha|s_i\rangle
 \end{equation}
is not only well defined, but also independent from the 
regularization of $D$.  This fact allows us to write our expansion 
directly in diff-invariant form as 
 \begin{eqnarray}
\langle s_f|U(T)|s_i\rangle &=&
\langle s_f|s_i\rangle \n \\ &&
+ (-iT) \left(
        \sum_{\alpha\in[s_i]} A_\alpha(s_i)\
        \langle s_f |D_\alpha|s_i\rangle +
        \sum_{\alpha\in[s_f]}
         A_\alpha(s_f)\
        \langle s_f |
D^\dagger_\alpha|s_i\rangle \right)
\n \\ && + {(-iT)^2\over 2!}
   \sum_{\alpha\in[s_i]}   \sum_{\alpha\prime\in[s_1]}
  A_\alpha(s_i)\  A_{\alpha\prime}(s_1)
 \langle s_f | D_{\alpha\prime}|s_1\rangle
\langle s_1 |\hat D_\alpha|s_i\rangle
\n \\
 & &  + \ \ \  \dots \ .
\label{e}
\end{eqnarray}

This expression gives the 3-geometry to 3-geometry propagator of 
quantum general relativity as a series finite at every order.  Notice 
that the expansion is in power of $T$, or, equivalently, in inverse 
powers of the Planck length, because this must divide $T$ in order to 
recover physical dimensions. The utility of a perturbation expansion in 
inverse powers of $G$ has been advocated in quantum gravity by Isham, 
Teitelboim and others \cite{teit}. 
Intuitively, we can think that this quantity represents the 
probability amplitude that if we have a quantum state of the 
gravitational field (a quantum 3-geometry) $|s_i\rangle$ over a 
surface $\Sigma_i$, we will find the quantum 3-geometry $|s_f\rangle$ 
on the surface $\Sigma_f$ in a proper time $T$.

\section{Sum over surfaces}

Surfaces in spacetime provide a natural bookkeeping device for the terms
of the expansion (\ref{e}) in the same manner in which Feynman graphs
provide a bookkeeping device for conventional QFT perturbation 
expansion. This fact leads us to give a nice graphical interpretation 
to the expansion (\ref{e}).  

Consider the 4-d 
manifold ${\cal M}=[0,1]\times\Sigma$.  Denote the two connected 
components of the boundary of $\cal M$ as $\Sigma_i$ and $\Sigma_f$.  
We now associate a 2d colored surface $\sigma$ in $\cal M$
--defined up to 4d diffeomorphisms-- to 
each nonvanishing term of the sum in the right hand side of (\ref{e}).
We begin by drawing $s_i$ in $\Sigma_i$ and $s_f$ in $\Sigma_f$.  As 
$s_i$ and $s_f$ have no information about the actual location of the 
graph, location is chosen arbitrarily (that is, up to a diffeomorphism in 
$Dif\!f_0$).  The first term in (\ref{e}) (zero-th order in $T$) is 
nonvanishing only if $s_f=s_i$.  In this case, let us slide $s_i$
across $\cal M$ from $\Sigma_i$ to $\Sigma_f$, in such a way that it 
ends up over $s_f$. To the term of order zero we associate  
the surface $\sigma=s_i\times [0,1]$ swept by $s_i$. See Figure 2. 

\begin{figure} \centerline{\mbox{\epsfig{file=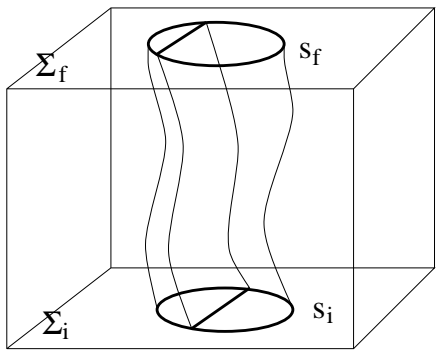}}} 
\caption{Surface corresponding to a term of order zero.} \end{figure}

Notice that this is possible because the two spin networks are in the 
same s-knot.  The surface we obtain is formed by 2d  
faces --submanifolds of $\cal M$-- joined along edges. The faces are 
swept by the spin network links, and the edges are swept by the spin 
network nodes.  We color every face with the color of the 
corresponding link of $s_i$, and every edge with the color of the 
corresponding node of $s_i$.

The surface associated to one of the summands of the second term 
(first order in $T$) in (\ref{e}) is then defined as follows.  In each 
summand, one of the nodes of $s_i$, say the node $i$, is altered by the 
operator $D$.  $s_f$ has two nodes more than $s_i$, say $i'$ and $i''$.  We 
begin by sliding $s_i$ into the manifold by an arbitrary finite amount, 
until a position, say, $s$.  Let $p$ be the point in which the node $i$ 
ends up.  Then we slide $s_f$ from $\Sigma_f$ through the manifold in such 
a way that it converges to $s$.  The three nodes $i, i'$ and $i''$ of $s_f$ 
converge all three to $p$.  We obtain a surface $\sigma$, bounded by 
$s_i$ and $s_f$ formed by faces that meet along edges; four of 
these edges meet at the point $p$.  We call $p$ a {\it vertex\/} of the 
surface $\sigma$.  At the vertex $p$, $\sigma$ branches.  Notice that 
four edges and six faces meet in $p$. See Figure 3.

\begin{figure} \centerline {\mbox{\epsfig{file=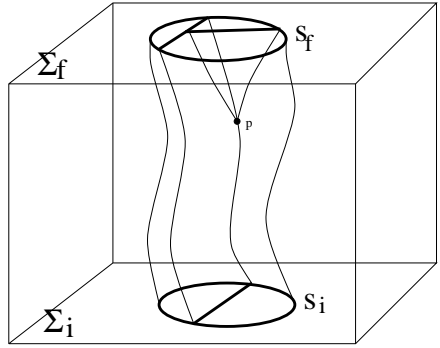}}} 
\caption{Surface
corresponding to a first order term.} \end{figure}

We can imagine $\cal M$ as a spacetime and $s_i$ as evolving continuously 
in a coordinate $t$ from $S_i$ to $S_f$.  At the spacetime event $p$, the 
spin network branches: the node $i$ generates the two new nodes $i'$ and 
$i''$, which are born at $i$ and then move away.  A new face, spanned by 
the new edge that joins $i'$ and $i''$, is born in $p$.  The branching 
represents the elementary vertex of the theory, and is represented in 
Figure 4.

\begin{figure} \centerline{\mbox{\epsfig{file=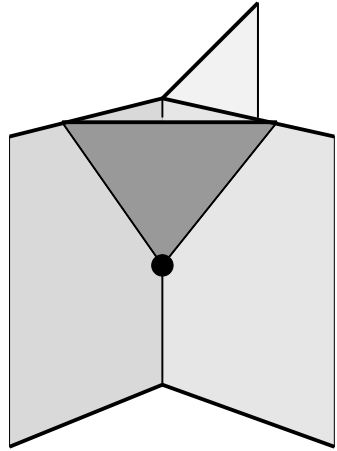}}} \caption{The
elementary vertex.} \end{figure}

The generalization of this construction to higher terms is immediate.  A 
term of order $n$ in $T$ corresponds to a surface $\sigma$ with $n$ vertices.  
The (time) order in which the $n$ $D$-operators act determines an ordering 
for the vertices.  An example of a term of order two is given in 
Figure 5.  It represents the transition from the s-knot with two trivalent 
nodes connected by three links colored (3, 5, 7), to the s-knot with the 
same graph, but colored (3, 6, 8).  The intermediate step is the s-knot 
$s_1$, with four nodes.

\begin{figure} \centerline{\mbox{\epsfig{file=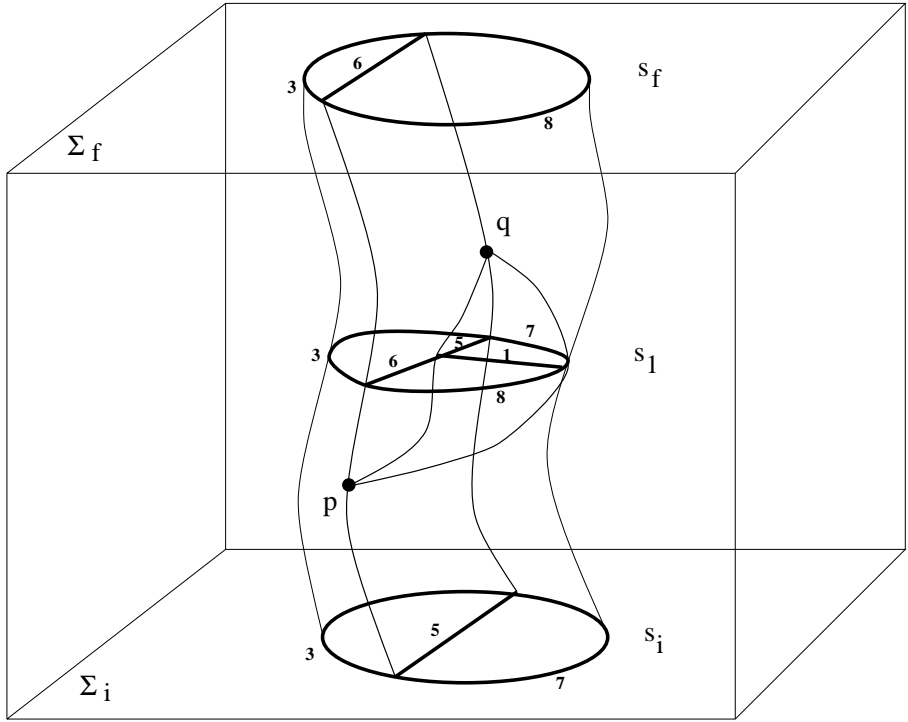}}}
 \caption{A term of second order.}
\end{figure}

In the construction we have described, each vertex has 
4 adjacent edges.  Some of these edges are generated by
the nodes of the incoming s-knot (the one at the
right of the operator $\hat D$ corresponding to that
vertex) and some by the outgoing s-knot. At each node, 
we denote the first ones as ``past'' edges and the second 
ones as ``future'' edges. Thus, each edge emerges as a 
future edge from one vertex, or from the initial hypersurface, 
and ends as a past edge in another vertex vertex, or in the
final hypersurface. This defines a partial ordering of 
the vertices of each surface. 

A short reflection will convince the reader that all the surfaces 
that we obtain satisfy the following property. Each face 
has the topology of a disk, and the ordered set vertices around a single 
face has at most one local maximum and most one 
local minimum. We say that a colored surface $\sigma$ is 
``well-ordered'' if all its faces satisfy this property. 

Now, observe the following.  (i) The colored surface $\sigma$ 
(with the vertices ordered) contains all the information needed to 
reconstruct the corresponding term in the expansion (\ref{e}).  In fact, 
the factors  $A_\alpha(s)$ depend only on the coloring of the surface.  
(ii) Any well-ordered branched colored surface $\sigma$, with colorings 
satisfying Clebsch-Gordan conditions at the edges can be obtained from a 
term in (\ref{e}).  (iii)  Two surfaces correspond to the same term if 
and only if there is a 4-d diffeomorphism that sends one into the other.  

These facts allow us to 
rewrite the expansion (\ref{e}) as a sum over diffeomorphic inequivalent 
well-ordered surfaces $\sigma$ bounded by $s_i$ and $s_f$.  Therefore we can
write the propagator (see (\ref{ptp})) as a sum of terms labeled by
topologically inequivalent branched well-ordered colored surfaces\footnote{
For a general description of such surfaces and their properties, see for 
instance \cite{cs} and references therein.} $s$ bounded by initial and 
final state: 
 \begin{equation}
P(s_f, s_i;\, T)  =
\sum_{\stackrel{\scriptstyle\sigma}{\scriptstyle\partial\sigma=s_i\cup s_f}}
 {\cal A}[\sigma](T)
\label{uno}. 
 \end{equation}
The weight ${\cal A}[\sigma](T)$ of the surface $\sigma$ is given 
by a product over the $n(\sigma)$ vertices of $\sigma$:
 \begin{equation} {\cal A}[\sigma](T)= 
{(\mbox{\rm - }\!i\;T)^{n(\sigma)}\over n(\sigma)!}\ 
\prod_{v\in[\sigma]} A_v(\sigma). 
\label{due}
\end{equation} 
The contribution $A_v(\sigma)$ of each vertex is 
given by the coefficients of the Hamiltonian constraint defined in 
equation (\ref{h4})
 \begin{equation} A_v(\sigma) = A_\alpha(s).
\label{tre}
\end{equation} 
 (The nonvanishing matrix elements $\langle s' |D_\alpha |s\rangle$ have 
value one.) 

Can we attribute a physical meaning to the surfaces that enter the sum?
The answer is yes.  There is a natural way of interpreting a branched
colored surface $\sigma$ as a discrete (``quantum'') geometry. This
geometrical interpretation was proposed in \cite{mike2} in a slightly
different context; it holds in the worldsheet formulation of the
simplicial model of GR \cite{mike}.   First of all,
consider a triangulation $\cal T$ of the manifold $\cal M$, and assume
that the surface $\sigma$ sits over the dual two-skeleton of the
triangulation. As we shall see in Appendix B, this is the natural way of
viewing the surfaces $\sigma$.  Let a triangle (two-cell) $S$ of the
triangulation $\cal T$ be punctured by the face $f$ (say with color
$p$) of $\sigma$ in a point.  Now, recall that according to canonical loop
quantum gravity the colors of the spin networks are quanta of area: the
area of a surface $S$ pierced by a single link with color $p=2j$ is
\cite{discr} 
\begin{equation}
        A(S) = 16\pi\hbar G\ \sqrt{j(j+1)}.  
\label{area} 
\end{equation}
In the spacetime picture, a link sweeps a 2d face $f$, which intersects
$S$ at a point.  It is natural to suppose that the area of any spacetime
2-surface $S$ is similarly determined by the coloring of the worldsheet.
For instance, we may consider a three-dimensional hypersurface $\Sigma$
that contains $S$, view the intersection between
$\Sigma$ and the colored surface $\sigma$ as the ``instantaneous position
of the spin-network state on the ADM time $\Sigma$'', and assume that the
results of the canonical theory can be applied. If we make this
assumption, then we can say that the area of $S$ is $A(S)$ given in
(\ref{area}).  Therefore, a surface $\sigma$ assigns a (possibly 
vanishing) area to 
each triangle of the triangulation $\cal T$.  But fixing the areas of the
triangles of a four dimensional triangulation is equivalent to fixing a
discretized 4-geometry. Assigning areas is analogous to assigning
the lengths of the links of the triangulation as in Regge
calculus\footnote{There is a difference: in order to define a geometry,
the lengths of the links must satisfy certain inequalities. The areas of
the triangles must satisfy certain inequalities, as well as some
equalities among them. Namely they are not all independent.} 
Thus, a surface $\sigma$ defines a discretized
4-geometry.  The idea that areas of triangles could be variables
more suitable than lengths of links in 4 dimensions was considered in 
\cite{pri,simp}. Finally, more in general, we can say that a natural 
geometrical interpretation of the colors associated to the faces
is the following: if a face has color $2j$, it contributes a quantum of
area $16\pi\hbar G\ \sqrt{j(j+1)}$ to the area of each spacetime 2-surface
$S$ at each point where it pierces $S$.  

This geometrical interpretation is ``natural'', but not necessarily
correct.  In particular, the relation between the proper time $T$, and the
spacetime geometry defined by the colors of the worldsheet is not clear.
This relation should be investigated before taking the geometrical
interpretation too seriously.

\subsection{Reconstruction of the $A_\alpha(S)$ coefficient from
surface data} 

The coefficients $A_\alpha(s)$ can be reconstructed directly from the colored
surface as follows.  Let a vertex $v$ have $n_i$ past edges and $n_f$
future edges.  $A_v(\sigma)$ is non vanishing only if $n_i=1$ and $n_f=3$
or if $n_i=3$ and $n_f=1$.  In this case, $A_\alpha(s)$ is determined by
the matrix elements of the hamiltonian constraint. 

It is very instructive to give an explicit construction of $A_v(\sigma)$.  
Consider a 4d neighborhood $B$ of the vertex $v$.  Consider the 3d 
boundary $\partial B$ of $B$.  Let $S_v$ be the intersection between $\sigma$ 
and $\partial B$.  A short reflection will convince the reader that $S_v$ 
is a colored graph in the 3-d space $\partial B$, having $4=n_i+n_f$ nodes 
(that satisfy Clebsh Gordan relations), corresponding to the intersections 
between the 4 edges emerging from $p$ and $\partial B$.  See Figure 6.

\begin{figure} \label{slicing_relativity}
\centerline{\mbox{\epsfig{file=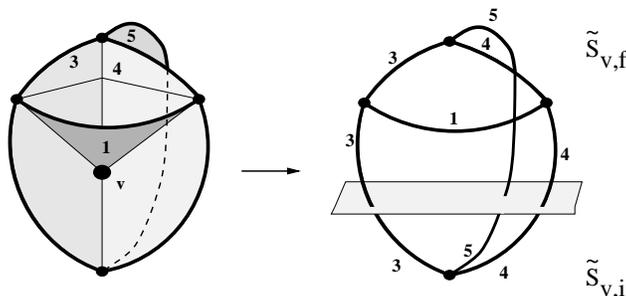}}} \caption{The
construction of the spin network $S_v$ from the intersection of the surface 
with the boundary of a 4-sphere surrounding the vertex.  The spin network 
is then cut into its past and future components $\tilde S_{v,i},
\tilde S_{v,f}$.} \end{figure}

Now, cut all the links of $S_v$ that go from a past node to a future node.  
This procedure breaks $S_v$ into two spin networks with (equal) open ends, 
which we denote as $\tilde S_{v,i}$ and $\tilde S_{v,f}$. (More precisely,
these are s-knots, because they are determined only up to
diffeomorphisms.)  The value of the vertex is given by the matrix elements
of the Hamiltonian constraint between these two spin networks.  Namely, 
\begin{equation} 
A_v(\sigma)=\langle\tilde S_{v,f}|C[1]|\tilde S_{v,i}\rangle
\label{Av};
 \end{equation}
see (\ref{tilde}).  This expression gives $A_v(\sigma)$ as a function of 
the colorings of the edges and faces adjacent to the vertex $p$. This 
function is universal, and characterizes general relativity, in the same 
manner in which the Feynman vertex factor characterizes a QFT.  We 
compute the vertex function $A_v(S)$ explicitly for the simplest case in 
Appendix B.  It turns out to be expressed in terms of $SU(2)$ n-$j$ 
symbols of the colorings.

Notice that it is the locality of the hamiltonian constraint that allows the 
sum over surfaces construction.  This is a peculiar form of 
locality, quite different from conventional QFT locality: 
The action is not local with respect to a background structure, but with 
respect to the spin networks themselves.

Equations (\ref{uno}-\ref{Av}) provide a definition of the proper time 
propagator of quantum general relativity as a topological sum over branched 
colored surfaces. They represent our main result. 

\section{Crossing symmetry}

Above, we have considered a {\it reformulation\/} of loop quantum gravity
as of a sum over surfaces. Here we propose a {\it modification\/} of the
theory, suggested by the reformulation. 

The value of the vertex $A_v(\sigma)$ that we have computed in the last
section depends on two inputs. First, on the coloring of the edges and
faces adjacent to the vertex $v$.  Second, on the distinction between
``past'' and ``future'' edges, namely on the way the
vertex is located and oriented within the surface $\sigma$. 
We suspect that the appearance of this orientation dependence is a sign
that something has got wrong in the definition of the theory. 
The action of GR is local and 4-d diff invariant, and therefore
the action of the 4-geometry of a small region ("the vertex") is
independent of how this region is sliced by equal time 
slices.\footnote{We are dealing with the
Euclidean theory, so there is no light-cone structure that defines local
notions of past and future.}

Thus, we propose a modification of the theory in which the orientation
dependence is removed.  We say, in general, that in a theory
defined by a sum over branched colored surfaces, with 
weights given by products of vertex factors, the vertex is ``crossing 
symmetry'' if its value depends on the adjacent colorings
only, and not on the distinction between past and future edges. 
BF theory \cite{CY} and simplicial GR \cite{mike2,mike} are theories
of this kind, and have crossing symmetric vertices.  In section V.A., we 
study the modification of the geometry of the vertex required to make 
it crossing symmetric.  

We then say that a hamiltonian constraint operator $H$ has crossing 
symmetry if it defines a crossing-symmetric vertex via eq.(\ref{Av}).  The 
modification of the vertex that we consider in section V.A might be 
obtained from a different factor ordering of the hamiltonian constraint,
and, as we shall see below, is strictly related 4d diff-invariance.  Thus,
here we are exploring the idea that 4d diff-invariance might fix residual 
factor ordering ambiguities. Of course, it should not be surprising that a
spacetime formalism could simplify the discussion of 4d diff-invariance, a
notoriously tricky issue in the hamiltonian framework. 

Let us make clear that we present crossing symmetry only as a 
proposal to be explored. 
We do not have a rigorous derivation of  crossing symmetry
from first principles, but only a heuristic plausibility argument, which
we better detail below. 

Consider the path integral that formally defines $U(T)$ in a
proper time gauge, namely in a gauge in which the lapse is spatially
constant. Consider a 4-metric $g$, that contributes to this path
integral and a small spacetime region $R$, and let $g_R$ be the
restriction of $g$ to $R$. The region $R$ is sliced by the proper time
slicing. Let $A_R(g)$ be the exponential of the action of this region. If
the region is small enough, we can think of $A_R(g)$ as the matrix element
of the evolution operator between ``before $R$'' and ``after $R$'', where
``before'' and ``after'' are determined by the proper time slicing, and
thus identify $A_R(g)$ with the vertex $A_v(\sigma)$. Now
consider a different 4-metric $g'$ in the
integral, containing a region $R'$, such that $g'_{R'}$ is isometric
to $g_R$, but sliced in a different manner by the proper time
slicing (the reader will easily convince himself that such a metric exists
in general). Since the action is local and 4-d diff invariant, the
contribution of $g_R$ to the sum must be equal to the contribution of
$g'_{R'}$, namely $A_R(g)=A_{R'}(g')$.  This implies that 
the matrix elements of the proper time hamiltonian between ``before'' and
``after'' according to one slicing of $R$ ought to be the same as the
matrix elements of between ``before'' and ``after'' according any other
slicing. In other words, the matrix elements should be invariant under a
4d rotation of $R$ that changes what is before and what is after. If we
require the same to hold in our sum over surfaces, we obtain the
requirement that vertices be crossing symmetric. 

This discussion shows that there is a relation between 4d
diff invariance and crossing symmetry, because a 4d diffeomorphism
``rotates'' the vertex in 4d.  
Recall that the 4d diff invariance of the classical theory is
expressed by the Poisson brackets
 \begin{equation}
\left\{ C[N],\ C[M]\right\} = C[N\vec\partial M-M\vec\partial N]. 
\label{pobr}
 \end{equation}
One of the hard problems of the hamiltonian quantization program is 
to define a quantization of the hamiltonian constraint yielding a 
4d diffeo invariant quantum theory. In particular, implementation of 4d
diff invariance is presumably the missing ingredient for fixing
quantization ambiguities of the hamiltonian constraint. Recall that the
ambiguity in the definition of $C[N]$ was fixed in \cite{hamiltonian} and
\cite{Thiemann} to a large extent arbitrarily.  Full implementation of the
quantum version of (\ref{pobr}) should ensure 4d diff invariance, but has
proven hard to realize.  We are therefore lead to the suggestion that we 
can cure the slicing dependence by taking advantage of the remaining 
operator ordering ambiguity, and at the same time, cure the excessive
ordering ambiguity by imposing slicing independence. 
In other words, we can {\it impose some form of 4d diff-invariance
requirement in order to reduce quantization ambiguity}. Here we are
suggesting that in a covariant formalism crossing symmetry
might be the key for implementing 4d diff invariance. 

\subsection{Vertices with crossing symmetry}

Let us investigate meaning and consequences of requiring crossing 
symmetry.  
First, we should require that different ways of cutting $S_v$ with 
three 
nodes on one side and one node on the other yield the same $A_v(\sigma)$.  
This leads to conditions on the $A_\alpha(S)$ coefficients, that will be 
studied elsewhere.  A more interesting case is the following.  First, let 
us help intuition by redrawing the elementary vertex of the theory (Figure 
4) in a more symmetric way.  This is done in Figure 7.  (For simplicity, we 
restrict the following analysis to trivalent nodes.)
 
\begin{figure} \centerline{\mbox{\epsfig{file=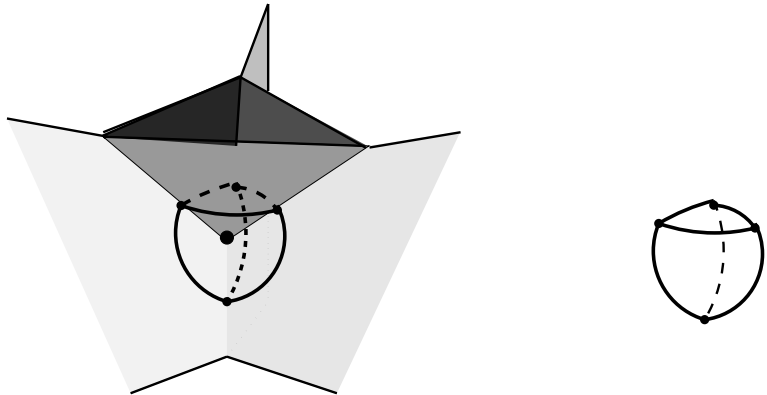}}} \caption{The
elementary vertex and its associated spin network $S_v$.} 
\end{figure}

There are five topologically inequivalent ways of cutting $S_v$, giving, 
respectively $(n_i, n_f)$ (number of initial and final nodes) equal to 
(0,4), (1,3), (2,2), (3,1) and (4,0).  The last two are the time reversal 
of the first two, leaving three genuinely independent cases.  In Figure 8, we 
show the possible cuts, and the corresponding spin networks transitions in 
the hamiltonian picture.  Time reversed cuts give just the opposite 
transitions.
 \begin{figure}
\centerline{\mbox{\epsfig{file=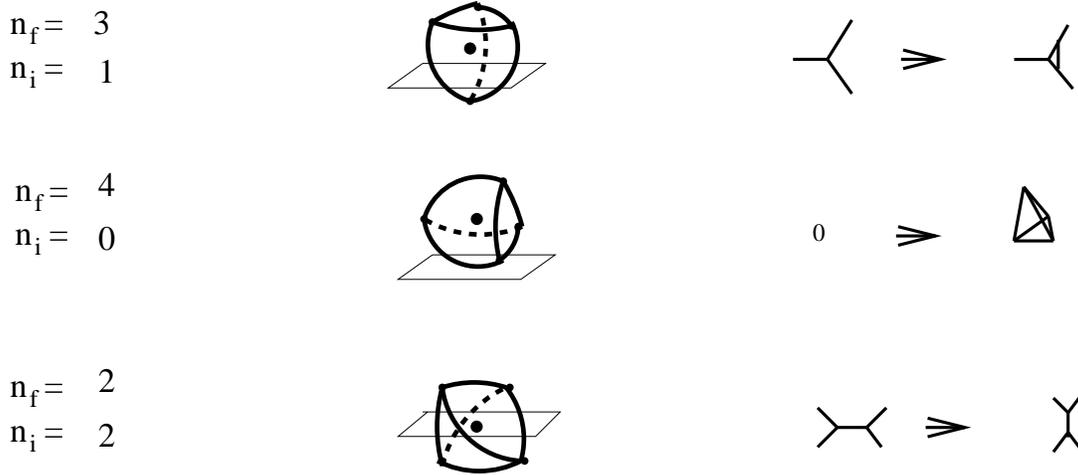}}}
  \caption{The (1, 3), (0, 4) and (2, 2) cuts of the elementary vertex, 
  and, in the last column, the corresponding spin network transitions in 
  the hamiltonian picture.  (For (3,1) and (4,0), look at (1,3) and (0,4) 
  upside down, and reverse the arrow of the transition.)} \end{figure} 

Case (1,3) is the one described in the previous section (Figure 6).  
Crossing symmetry requires that the hamiltonian constraint generate 
the transitions (0,4) and (2,2) -- described in the last column of Figure 
8-- as well, {\it with the same amplitude}.

Consider these two new transitions.  We begin with (0,4) (second 
line in Figure 8).  This transition represents a matrix element of a 
hamiltonian that creates a ``small'' tetrahedron from the state with no 
loops.  The fact that 4d invariance requires the presence of such 
``birth'' terms has already been argued, on general grounds, in
\cite{mike3}.  In terms of surfaces, the term looks as in Figure 9.
 \begin{figure}
 \centerline{\mbox{\epsfig{file=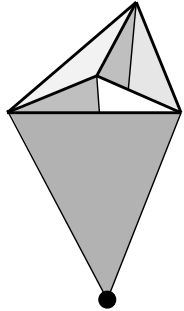}}}
 \caption{The (0,4) transition: creation of a tetrahedron.}
 \end{figure}
This is {\it the very same surface as in Figure 4, and in Figure 7}; but 
drawn with a different orientation in ``spacetime''.  With this 
orientation, it describes a tetrahedral spin network emerging from 
nothing.

Can such a term originate from an ordering of the hamiltonian 
constraint?  Surprisingly, the answer is positive.  We sketch here an 
hand waving argument.  The regularized hamiltonian constraint ($FEE$) 
is formed two parts: s ``small loop'' that corresponds to the 
classical curvature ($F$) term; and the term that ``grasps'': the two 
hands of the $T^2$ operator in \cite{loops,hamiltonian} (or the Volume 
operator in \cite{Thiemann}),  corresponding to the triads $EE$
(or the triads multiplied by a suitable density factor).  
Traditionally, the order chosen is $FEE$: the ``small loops'' is added 
{\it after} the grasping.  Reverse this order, choosing $EEF$, and 
have the small loop being inserted first.  Then the hamiltonian 
constraint has non-vanishing action on the vacuum as well, because the 
grasping term can grasp the ``small loop''.  In particular, this may 
create a ``small'' tetrahedron.  For instance, in the construction in 
\cite{loops,hamiltonian}, the $T^2$ can grasp itself, producing, precisely, 
a tetrahedron.  See Figure 10.

 \begin{figure}
 \centerline{\mbox{\epsfig{file=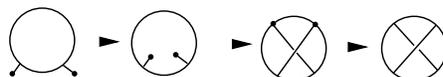}}}
 \caption{Creation of a tetrahedron from the self grasping of $T^2$.}
 \end{figure}

The (2,2) term (third line of Figure 8) gives a rearranging of two 
nodes. The corresponding surface looks as in Figure 11.

 \begin{figure}
 \centerline{\mbox{\epsfig{file=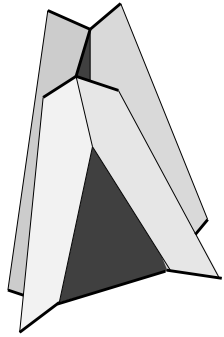}}}
 \caption{The (2,2) vertex.}
 \end{figure}

Again, this is just a different orientation of the same elementary 
vertex. 

Now, we could search for an ordering of $C[N]$ yielding a vertex having 
crossing symmetry.  But this task is superfluous since we already know what 
we should obtain.  We can directly {\it postulate\/} that the hamiltonian 
constraint yield crossing symmetry, and deduce the amplitudes of 
the (0,4) and (2,2) matrix elements from the value of the (1,3) 
vertex.\footnote{Notice that the symmetrization of the hamiltonian 
constraint given in (\ref{h2}) can be seen as a first step in this 
covariantisation of the operator: it is equivalent to the postulate that 
the (3,1) cut has the same value as the (1,3) cut.} 

There is a physical motivation that supports the above argument.  It 
has been observed \cite{bernie} that the orderings of the hamiltonian 
constraint studied so far generate a dynamical evolution that appear 
to be excessively ``local''.  They preserve the general structure of 
the network on which they act, simply ``dressing'' nodes.  These 
difficulties have been recently detailed in Ref.\cite{lee}, where it 
is argued that no long range interaction is likely to emerge from a 
hamiltonian with these features.  As pointed out by Thiemann, the 
argument is far from conclusive, because it contains a jump from the 
non-physical (gauge) coordinate evolution to the physical one, and 
this jump may be un ungranted.  In any case, adding the new vertices 
(2,2) and (0,4) would cure these potential difficulties.  

Also, we note that one of the consequences of adding the new vertices is 
that the faces of the surfaces in the sum do not need anymore to be 
(topologically) two-disks, as followed from the original Hamiltonian 
expansion, and there is no sense in the requirement of the surfaces being 
well-ordered. 

Finally, notice that one could also search for the form of the vertices 
from general a priori requirements.  An arbitrary crossing symmetric vertex 
is obtained by replacing (\ref{Av}) with a function $A(S_v)$ of the spin 
network $S_v$ associated to the vertex 
\begin{equation} 
A_v(\sigma) \equiv A(S_v).
\end{equation}
For instance, for trivalent nodes, $S_v$ is a tetrahedron: the 
function $A(\ \ )$ must respect tetrahedral invariance. Notice that there 
are not many functions with these features.  A natural choice is
\begin{equation}
	A(S_v) = {Tet\!\left[\begin{array}{ccc}
           a   &  b  & c \\
            d & e & f
        \end{array}\right]
      }
\end{equation}
where $a-f$ are the colors of the links of $S_v$ and $ \ Tet\ $ is 
the totally symmetric form of the 6-$j$ symbols (\cite{Kauffman94} 
and \cite{dpr}).  We think that a theory defined in this way is worth 
exploring.

\section{Conclusion}

Our main result is contained in the equation
\begin{equation}
P(s_f,s_i;\; T) =
\sum_{\stackrel{\scriptstyle\sigma}{\scriptstyle\partial\sigma=s_i\cup s_f}}
{(\mbox{\rm - }\!i\;T)^{n(\sigma)}\over n(\sigma)!}\ 
\prod_{v\in[\sigma]} A_v(\sigma). 
\label{expansion2}
\end{equation}
[see (\ref{uno}-\ref{due})], which expresses the dynamics of 
quantum general relativity in terms of a sum over surfaces $\sigma$.  

More precisely, the proper time propagator of quantum GR can be expressed 
in terms of a sum over topologically inequivalent branched colored 
surfaces, bounded by the initial and final s-knots.  The contribution 
of each surface to the sum is the product of one factor per each vertex 
(branching point) of the surface.  The contribution a each vertex is a 
simple $SU(2)$-invariant function $A_{v}(\sigma)$ of the colors of the
faces and edges adjacent to the vertex.  This function characterizes the
quantum theory in the same manner in which the Feynman graph vertices
characterize a quantum field theory.  The vertex $A_{v}(\sigma)$ of
general relativity is given by a product of Wigner $3n-j$ symbols
\cite{rrr}.

The essential property of the expansion (\ref{expansion2}) is that it 
is finite order by order, and  explicitly computable.  This finiteness is
intriguing.  In order to calculate physical quantities, we must have the 
proper time propagator for multifingered proper times and we 
must integrate over the multifingered proper time.  We expect that the
integration  could yield finite results if performed over  
expectation values of appropriate physical quantities.  Work is in 
progress in this direction, and will be reported elsewhere.

We close with some comments. 
\begin{itemize} 

\item Our construction is strongly reminiscent of discretized quantum 
gravity on a lattice \cite{loll,lattice,Iwasaki,zapata}, 
particularly in its simplicial formulations \cite{mike,simp}.  
It is shown in  \cite{mike} 
that one can discretize general relativity over a simplicial lattice, and 
express the gravitational degrees of freedom as colored branched surfaces 
over the (dual) two-skeleton.\footnote{Surfaces seem to be playing an 
increasing role as a way to capture the gravitational field degrees of 
freedom.  See for instance \cite{ted}.}
Even more remarkably, the partition function is given in the discretized
case by a construction very similar to that given here: the contribution
of a vertex is determined by the intersection between the boundary of a
4-simplex around the branching point of the surfaces, and the surface. 
This defines a spin network $S_v$, which, in the discretized case, can be
any subgraph of the 1-skeleton of a 4-simplex.  Therefore vertices have up
to 5 edges and 10 faces (see Appendix B) in the discretized case.  In this
paper, nonvanishing vertices have 4 edges and an arbitrary number of
faces.  Thus, the simplicial construction corresponds to a cut of the sum
(\ref{uno}) in two respects: the maximum number of vertices is fixed by
the triangulation, and vertices have 10 faces at most. 

\item One can view the sum over surfaces defined here as a version of
Hawkings' integral over 4-geometries.  Indeed, a colored two surface 
defines a discrete 4-geometry.  The integral is replaced here by a 
sum, and explicit computation can be performed.  Presumably, the 
construction then be used to define a number of related theoretical 
tools as partition function, Hartle-Hawking state, and similar 
\cite{hh}.

\item Each individual term in the expansion (\ref{e}) is finite.  
Divergences can arise in summing the series, and in integrating over 
proper time.

\item The similarity with the formulation of string theory as a path 
integral over worldsheets is tantalizing. On this, see the discussion in 
\cite{baez}.  The {\it  dynamics\/} is different.  In string 
theory, the contribution of each surface to the sum is given by the area 
of the surface, and therefore it depends on a fixed background metric on the 
manifold.  Here, on the contrary, the  contribution of each surface 
depends only on the (coloring and) topology of  the surface.  Thus, 
quantum GR resembles a ``background independent''  version of string 
theory.  The techniques developed here could perhaps have  relevance for 
connecting loop quantum gravity with string theory \cite{baez,smolin} -- 
or for the construction of a non-perturbative background independent 
formulation of string theory.

\end{itemize}

\vskip1cm

\centerline{------------------------------------------}

We are particularly indebted with Jim Hartle, Gary Horowitz and 
John Baez for several discussions during which various aspects of 
this work emerged and were clarified.  We thank Don Marolf for key 
suggestions in an early stage of the work, Bernie Br\"ugmann for 
criticisms and suggestions, Thomas Thiemann for a detailed and valuable 
critical reading of the manuscript, Roberto DePietri and Roumen Borissov
for pointing out and correcting several errors, and Lee Smolin, Seth 
Major, Jerzy Lewandowski, Louis Crane, John Baker and Abhay Ashtekar 
for suggestions and discussions.  This work was partially supported by 
the NSF Grants PHY-5-3840400, PHY-9515506, PHY95-14240 and the Eberly 
research fund of PSU. CR thanks the Albert Einstein Institute in Potsdam, 
and the Physics Department of the University of Roma ``La Sapienza'' 
for hospitality during the preparation of this work. 

\appendix

\section{Terminology}

To help the reader, we collect here a list of terms employed.
\begin{description}
	\item[Node:\ \ ] \, Point in 3d space where the links of a spin network 
    meet.

    \item[Link:\ \ ] \ \ Line in 3d space connecting two nodes of a spin 
network.

    \item[Face:\ \ ] \ \  Surface in 4d spacetime (swept by a link).

    \item[Edge:\ \ ] \ \, Line in 4d spacetime where several faces meet 
(swept by a node). 

    \item[Vertex:\ \ ] Point in 4d spacetime where several edges meet. 

\end{description}

A spin network is formed by nodes and links.  A branched surface is 
formed by faces, edges, and vertices.  
For the branched surfaces that live on the 
2-skeleton of the dual triangulation of the manifold in simplicial BF 
theory, faces, edges, and vertices live on 2-, 1- and 
0-cells respectively of the cellular decomposition dual to the simplicial 
triangulation. They are therefore associated to 4-, 3- and 2-simplices 
of the triangulation, respectively.  Therefore a vertex corresponds to a 
4-simplex, an edge to a tetrahedron, and a face to (its dual) triangle.

\section{Comparison with the Ooguri-Crane-Yetter 4d TQFT}

The structure of quantum general relativity in the form presented in 
this paper is surprisingly similar to the Ooguri-Crane-Yetter (OCY) four
dimensional topological quantum field theory \cite{ooguri,CY}  
a rigorously defined simplicial lattice version of four
dimensional $SU(2)$ BF theory. 

More specifically, our expression for the proper time propagator $U(T)$ of
GR as a sum over worldsheets resembles in many ways the worldsheet sum 
\cite{mike2,Kauffman94}\footnote{%
The worldsheet sum of \cite{Kauffman94} is actually for the 3d 
Ponzano-Regge-Turaev-Viro (PRTV) model, but is easily extended to the OCY 
model. Iwasaki \cite{Iwasaki} has proposed an interesting alternative, a 
closely related worldsheet formulation of the PRTV model which is also 
easily extended to the OCY model. See also \cite{barret}}
for the projector on physical states, $U_{OCY}$, of the OCY model.

The OCY model is a 4d generalization of 
the Ponzano-Regge-Turaev-Viro (PRTV) model \cite{PR,TV}, which, in turn, can 
be seen as a quantization of 3d GR, or a quantization of 3d Chern-Simon 
theory.  In \cite{pri} it was shown that the PRTV 
model is a theory of the dynamics of spin networks (loops in the terminology
of \cite{pri}) having the same 
physical interpretation as the spin network basis states in continuum 3d GR.
Thus one might expect a similarity between the kinematic of the OCY model
and loop quantized GR \cite{pri}. On the other hand, the 4d OCY model like
the 3d PRTV model, but unlike 4d GR has no local degrees of freedom, so
one also expects large differences between the theories.

In this section, we sketch the OCY theory, outline a construction of 
the worldsheet sum for the partition function $Z_BF$ of the OCY model
along the lines of \cite{mike2},\footnote{
The construction outlined here is a sort of baby version of that for
simplicial GR in \cite{mike3}.}
and
discuss its similarities and differences with the formulation of quantum GR
presented here. We believe that this comparison helps illuminate the much 
debated issue of the relation between quantum gravity and TQFTs  \cite{jmp}.

We introduce here Ooguri's original version \cite{ooguri} the OCY model 
heuristically, as a discretization of BF theory without cosmological 
constant.  BF theory is given in terms of two 
fields, an $SU(2)$ connection $A^i$, with curvature $F^{i}$, and an 
$su(2)$-algebra valued 2-form $B^i$, by the action \cite{BF}
\begin{equation}
S_{BF}=\int B^i\wedge F^i.
\end{equation}

Before proceeding, it is interesting to note that conventional general 
relativity can be obtained from BF theory by simply adding a 
constraint term.  Indeed one can show that the theory
\begin{equation}
S_{GR}=\int B^i\wedge F^i + \phi_{ij}B^i \wedge B^j,
\end{equation}
where the Lagrange multiplier  $\phi_{ij}$ is traceless and 
symmetric, is equivalent to general relativity \cite{BFGR}.

Consider the partition function of the BF theory
\begin{equation}
Z_{BF}= \int [dA]\ [dB]\  e^{-i \int B^i \wedge F^i }.
\end{equation}
Integrating over $B$, we obtain
\begin{equation}
Z_{BF}= \int [dA] \ \delta[F],
\label{Z}
\end{equation}
namely an integral over flat $SU(2)$ connections.  Let us define a lattice 
version of this theory by fixing a simplicial decomposition of the 4d 
manifold. (See also \cite{zapata}.) 

Consider the dual of the simplicial decomposition.  There is one
element of this dual cellular decomposition that plays a central role
in the construction: the ``wedge''. Consider a dual-two-face $f$.  A
dual-2-face is a 2d polygon. It intersects a two-face of the
simplicial decomposition in a ``central'' point $o$. Its vertices are
centers of 4-simplices and its sides are lines connecting such
centers. Each of this sides, which
connects the centers of two simplices, crosses the tetrahedron that forms 
the boundary between the two simplices. Let $p$ be the crossing point.
Each point $p$ cuts one of the sides of the polygon $f$.  By drawing lines
connecting the the points $p$ to $O$, we divide the polygon $f$ in
quadrangles, called ``wedges''. A wedge is thus a 2d quadrangle that 
has four sides: two of these are 1d lines that join centers of 4-simplices
with (the center $p$ of) a bounding tetrahedron; these are denoted 3-4
flags. The other two lines join the center of a tetrahedron with the
center of a two-face. These are denoted 3-2 flags. See Figure 8.

  \begin{figure}
  \centerline{\mbox{\epsfig{file=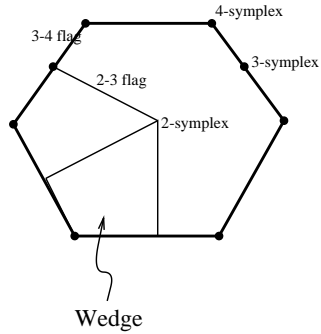,width=4cm}}}
  \caption{A dual-two-face and its decomposition in wedges.}
  \end{figure}

We choose to represent the connection by means of
group elements associated to 1d elements in the dual cellular
decomposition. More precisely, we associate a group element $U$ to each
2-3 flag (segment connecting the center of a 2-simplex
with the center of one of the tetrahedra surrounding it), and 
one group element $W$ to each 3-4 flag (segment
connecting the center of a tetrahedron with the center of an
adjacent 4-simplex).  These group elements can be thought as the
exponential of the connection along the segments. Each wedge 
$w$ is bounded by four of such segments (two of 
the 2-3 kind and two of the 3-4 kind); let $U_1(w),U_2(w), W_1(w), W_2(w)$ 
be the group elements associated to the segments that bound the wedge $w$.  
We can express the requirement that the connection is flat by requiring 
that the holonomy of the connection around each wedge is trivial.  Then a 
discretization of (\ref{Z}) is given by
\begin{equation}
Z_{BF}= \int [dU]\ [dW] \ \prod_w \delta[U_1(w)U_2(w)W_1(w)W_2(w)]
\label{Zd}
\end{equation}
where the delta function is the delta function of the unit on the 
$SU(2)$ group.  We can expand the delta function in characters. 
For each $w$, we have 
\begin{equation}
\delta[U_1(w)U_2(w)W_1(w)W_2(w)]=\sum_j (2j+1) 
Tr_j(U_1(w)U_2(w)W_1(w)W_2(w)),
\end{equation}
where $j$ labels the irreducible representations of 
$SU(2)$, and $Tr_j(U)$ 
is the trace of the group element $U$ in the representation $j$. 
Using this, we can rewrite (\ref{Zd}) as an integral over group elements 
$U$ and $W$ associated to segments and half integers $j$ associated to wedges:
\begin{equation}
Z_{BF} = \int [dU]\ [dW] \sum_{[j]} \prod_w 
(2j(w)+1) Tr_{j(w)}(U_1(w)U_2(w)W_1(w)W_2(w)). 
\label{Zd2}
\end{equation}
We can view the group elements $U$ and $W$ as the discrete version of the 
connection $A$, the $j$'s as a discrete version of the two form 
$B$ and the expression $Tr_{j(w)}(U_1(w)U_2(w)W_1(w)W_2(w))$ as the 
discrete version of the expression $\exp(-i B^i\wedge F^i)$. 

Next, let us perform the group integration in (\ref{Zd2}) explicitly.  By 
integrating over the group elements $U$ we force the colors of the wedges 
belonging to the same dual 2-cell to be equal.  By integrating over 
the group elements $W$ we force the $j$'s of the (four) dual 2-cells 
that join on a 1-cell to satisfy Clebsh-Gordan relation where they meet, 
leaving an extra degree of freedom $J$ associated to each such dual 
1-cell; $J$ runs over the independent couplings of four $SU(2)$ 
representations. Finally,  
we end up with numerical factors associated to the 0-cells of the dual 
triangulation (plus other factors associated to faces and edges, 
which we disregard here in order not to make the exposition to heavy). 
Such numerical factors turn out to be 15-$j$ symbols associated to the 
five $J$'s of the five 1-cells and the ten $j$'s of the ten 2-cells  
adjacent to each vertex.\footnote{
     A dual 0-cell is always adjacent to five 1-cells and ten 2-cells, 
     because it corresponds to a four-simplex of the original 
     simplicial decomposition, which is bounded by five tetrahedra and 
     ten faces.}.  
Performing these integrations explicitly is a simple and 
interesting exercise.  After these 
integration over the group elements, the theory is therefore reduced 
to a sum over colorings on the 2-cells and 1-cells, satisfying 
Clebsh-Gordan relations.  We can 
interpret a zero color as no surface at all, and identify the 2-cells
with faces and the 1-cells with edges of branched colored surfaces. 
Thus, we can write the partition function as a sum 
over branched colored surfaces living on the dual two skeleton of the 
simplicial triangulation.  We obtain 
\begin{equation} 
       Z_{BF} = \sum_\sigma A_{BF}[\sigma]
\end{equation}
where the contribution of each surface is (up to the face and edge 
factors we  have disregarded for simplicity) a product of vertices's 
factors
\begin{equation} 
       A_{BF}[\sigma] = \prod_{v} A_{BF, v}(\sigma);  
\end{equation}
the vertex factor is the 15-$j$ symbol of the colorings adjacent 
to the vertex.  

The similarity of this result with the construction in 
this paper, equations (\ref{uno},\ref{due},\ref{tre}), is striking.  In 
both cases, we have a sum over the same kind of branched colored 
surfaces, and the weight for each surface is the product of vertex 
factors, where vertex factors are simple $SU(2)$ invariant functions 
of the adjacent colorings.  Thus, the structure of quantum general 
relativity turns out to be extremely similar to the structure of a 
topological quantum field theory.  Of course there are differences, 
and these differences are crucial.  Let us examine them in detail. 
\begin{itemize}
	\item  First of all, the worldsheet amplitudes in the BF theory
	that we are considering are the amplitudes in the projector on
	physical states, while the GR worldsheet amplitudes of the 
	present paper are from the sum for the proper time propagator 
	$U(T)$, so we might be comparing apples and oranges. However,
	if we accept the not unreasonable hypothesis that $U(T)$ is a
	partial sum of terms in a sum over surfaces for $U$ in GR, we
	can compare the theories in a direct way. 
	\item  The vertex factor is different in the two 
	theories: in BF theory it is a 15-$j$ symbol, while in GR it is a 
	combination of 9-$j$ and 6-$j$ symbols. This difference depends on 
	the different dynamics of the two theories, and should be at the 
	root of the other differences.
	\item  In the case of BF theory there is a crucial theorem 
	holding: triangulation independence. Refining the triangulation does 
	not change the overall sum. This is the reason for which the theory 
	is topological, and is a consequence of the fact that the classical 
	theory has no local degrees of freedom. In GR, nothing similar 
	holds, because GR has genuine local ({\it although non-localized!\/}) 
	degrees of freedom. Therefore there is no reason to expect anything 
	like triangulation independence for GR.
	\item  The ensemble of surfaces over which the sum is defined is 
	different in the two cases. In the BF case, we sum over surfaces over  
	a fixed triangulation. In the case of GR, we sum over all 
	topologically inequivalent surfaces, with an arbitrary number of 
        vertices. 
	Therefore, in the case of BF theory the 
	surfaces to be consider are finite in number. 
	In the case of GR we have to sum over 
	arbitrarily complicated surfaces, or, equivalently, sum over 
	arbitrarily fine triangulations of the manifold as well.  Notice that 
	this difference is a consequence of the 
	previous point, namely triangulation independence of BF. We 
	could average over arbitrarily fine triangulations in BF as well; but 
	this would not affect the result, because each triangulation 
	yields the same contribution as the coarsest one. Therefore:  
	{\it Diff invariance of the sum is implemented in two 
	different ways in BF and in GR, corresponding to the fact that BF is 
	topological, while GR is not: in BF, invariance is obtained thanks to 
	triangulation independence; in GR invariance is obtained by 
	summing over arbitrarily fine triangulations.}
	\item  The rigorous version of BF theory requires $SU(2)$ 
	to be replaced by quantum $SU(2)$. This can be seen simply as a 
	smart stratagem for regularizing the sum in an invariant manner, 
	yielding a finite result. Notice that in GR regularizing $SU(2)$ to 
	quantum $SU(2)$ would not guarantee an overall finite sum, because the 
	surfaces themselves are infinite in 
	number.  Thus, quantum GR does not admit a rigorous finite version 
	as quantum BF, at least as far as we can presently see, even if one 
	attempts to replace $SU(2)$ with quantum $SU(2)$ in GR \cite{seth}.
	\item The vertices of BF have always five edges and ten faces, 
	while vertices of GR have (at least with the 
        ordering considered so far) 
	four edges and an arbitrary number of faces. 
\end{itemize}
These points illuminate the difference between quantum 
GR and topological field theories.  Let us discuss this point in more 
detail. 

Both theories are invariant under diffeomorphism. 
However, diffeomorphism invariance does not imply 
that a quantum theory is topological in the sense of having a finite 
number of degrees of freedom.
We expect GR to have an infinite number of degrees of freedom. Thus
Atiyah's axioms for topological quantum field theory are 
likely to be suitable for quantum general relativity as well, if we 
drop the request that the Hilbert spaces attached to boundaries of the 
four-manifold be finite dimensional.

Finally, we may turn the comparison the other way around, and 
describe quantum BF theory in the language so far used for quantum GR. 
We can capture quantum BF theory in terms of its vertex.  A BF vertex 
has five edges and ten faces.  Assume that one of these edges comes 
from the past, and four go to the future (the other cases are given by 
crossing symmetry, that clearly holds in BF theory).  
A moment of reflection 
shows that the elementary vertex of BF theory ``opens up'' a 
four-valent intersection of a spin-network into a ``small'' 
tetrahedron:

\begin{figure} \centerline{\mbox{\epsfig{file=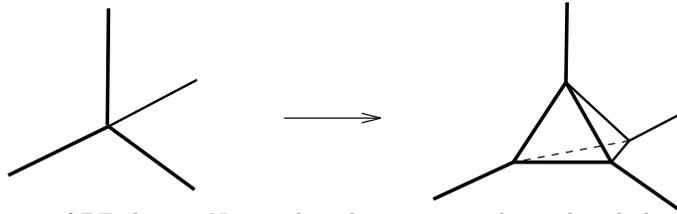}}} 
\caption{The elementary vertex of BF theory. Notice that there are 
5 nodes and 10 links, yielding 15 colors. The value of this vertex is the 
15-$j$ symbol of their 15 colors.} 
\end{figure}

The matrix element of the hamiltonian between these two (partial) spin 
networks is the 15-$j$ symbol of the 15 colors associated to the four 
links and the one node  of the incoming spin network, and the six other 
links and four other nodes of the newly created tetrahedron. It 
would be interesting to derive this hamiltonian from a hamiltonian 
loop quantization of BF theory.

\section{Diff invariant scalar product}

We work out here an example of diff-invariant 
scalar product between s-knot states \cite{gi}.  
(On the inner product between spin networks see 
\cite{inverse,Baez95a}, and \cite{dpr} which we follow here).  Let 
$s$ be the s-knot defined as follows.  $s$ has three 4-valent nodes, 
$i,j$ and $k$, and the six links
  \begin{equation} 
    (ki,2),\ (kj,2),\ (ki,4), \ (kj,4),\ (ij,3),\ (ij,5), 
  \end{equation} 
where each link is indicated by the two nodes it 
connects and its color.  Explicitly:

\begin{equation}
\begin{array}{c}
 \mbox{\epsfig{file=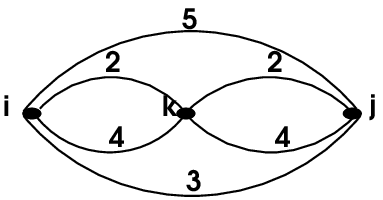}}
 \end{array}
\end{equation}
To specify the state, we have to give also the coloring of 
the nodes.  We choose an expansion of the nodes in a trivalent graph 
(in each node) by pairing the two links colored 2 and 4.  We have one
virtual link for every node (we denote them $e_i, e_j$ and $e_k$), 
which we assume to be colored as follows: $e_i: c,\ \  e_j:6$ and $ 
e_k: 2$, where $c$ (which can take the values 6, 4 and 2) will be specified 
later on
\begin{equation}
\begin{array}{c}
 \mbox{\epsfig{file=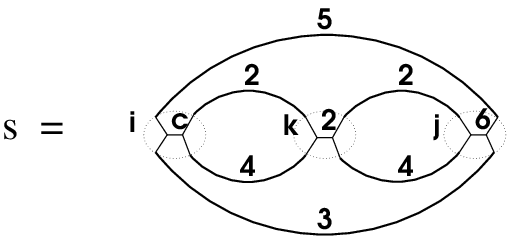}}
 \end{array}
\end{equation}
Next, let us define the s-knot $s^\prime$.  Let it be the same as above, but 
with a different coloring of the node $j$.  We expand $j$ by pairing $(kj, 
2)$ with $(ij, 5)$ and $(kj, 4)$ with $(ij, 3)$.  Let the internal link 
have color 3:
\begin{equation}
\begin{array}{c}
 \mbox{\epsfig{file=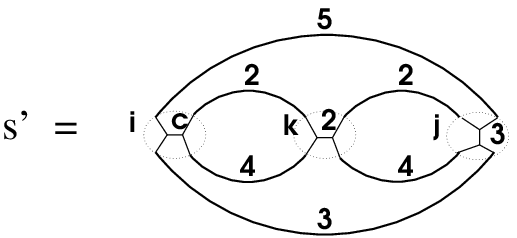}}
 \end{array}
\end{equation}
Let us compute the scalar product $\langle s|s^\prime\rangle$. First, 
we have to list the automorphisms of the spin network (taking links' 
colors, but not node's colors into account). There is only one non trivial 
automorphism $\alpha$: it exchanges $i$ and $j$. Thus, 
(\ref{product}) gives
\begin{equation}
\langle s|s^\prime\rangle = \langle S|S^\prime\rangle + 
\langle \alpha S|S^\prime\rangle 
\label{sp1}
\end{equation}
where $S\in s$ and $S^\prime\in s^\prime$ have been selected to 
have the same graph, with the same colored links. The only contribution to 
(\ref{sp1}) comes from the nodes. The node $k$ is the same in the 
two states and therefore gives no contribution (recall we have chosen 
normalized states). Thus, we have
\begin{equation}
\langle s|s^\prime\rangle = 
\langle S|S^\prime\rangle_i
\langle S|S^\prime\rangle_j
 +
\langle \alpha S|S^\prime\rangle_i
\langle \alpha S|S^\prime\rangle_j
\label{sp2}
\end{equation}
where we have indicated with $\langle\ |\ \rangle_i$ the scalar 
product restricted to the space of one node. Spin-network states with the 
same trivalent expansion are orthonormal. The change of basis is given  
\cite{dpr} by the recoupling theorem 
\begin{equation}
\begin{array}{c}\setlength{\unitlength}{1 pt}
\begin{picture}(50,40)
          \put( 0,0){$a$}\put( 0,30){$b$}
          \put(45,0){$d$}\put(45,30){$c$}
          \put(10,10){\line(1,1){10}}\put(10,30){\line(1,-1){10}}
          \put(30,20){\line(1,1){10}}\put(30,20){\line(1,-1){10}}
          \put(20,20){\line(1,0){10}}\put(22,25){$j$}
          \put(20,20){\circle*{3}}\put(30,20){\circle*{3}}
\end{picture}\end{array}
    = \sum_i  \left\{\begin{array}{ccc}
                      a  & b & i \\
                      c  & d & j
              \end{array}\right\}
\begin{array}{c}\setlength{\unitlength}{1 pt}
\begin{picture}(40,40)
      \put( 0,0){$a$}\put( 0,40){$b$}
      \put(35,0){$d$}\put(35,40){$c$}
      \put(10,10){\line(1,1){10}}\put(10,40){\line(1,-1){10}}
      \put(20,30){\line(1,1){10}}\put(20,20){\line(1,-1){10}}
      \put(20,20){\line(0,1){10}}\put(22,22){$i$}
      \put(20,20){\circle*{3}}\put(20,30){\circle*{3}}
\end{picture}\end{array}
\label{recoupling}
\end{equation}
where the quantities $\left\{\begin{array}{ccc}
a & b & i \\ c & d & j  \end{array}\right\}$ are $su(2)$
six-j symbols (normalized as in \cite{Kauffman94}).
This gives us immediately 
 \begin{eqnarray}
\langle S|S^\prime\rangle_i &=& 1 \\
\langle S|S^\prime\rangle_j &=& 
\left\{\begin{array}{ccc}
4 & 2 & 3 \\ 5 & 3 & 6  \end{array}\right\} \\
\langle \alpha S|S^\prime\rangle_i &=& \delta_{c6} \\
\langle \alpha S|S^\prime\rangle_j &=& \left\{\begin{array}{ccc}
4 & 2 & 3 \\ 5 & 3 & c  \end{array}\right\} 
\end{eqnarray}
Therefore 
\begin{equation}
\langle s|s^\prime\rangle =
\left\{\begin{array}{ccc}
4 & 2 & 3 \\ 5 & 3 & 6  \end{array}\right\}
+  \delta_{c6}\left\{\begin{array}{ccc}
4 & 2 & 3 \\ 5 & 3 & c  \end{array}\right\}.
\end{equation}
Thus, if $c$ = 6 there are two contributions to the scalar product, one 
from each of the two elements of the automorphism group of the spin network
and we have
\begin{equation}
\langle s|s^\prime\rangle =
2 \left\{\begin{array}{ccc}
4 & 2 & 3 \\ 5 & 3 & 6  \end{array}\right\} = {112\over 75}.
\end{equation}
While if $c=2$ or $c=4$
\begin{equation}
\langle s|s^\prime\rangle =
\left\{\begin{array}{ccc}
4 & 2 & 3 \\ 5 & 3 & 6  \end{array}\right\} = {56\over 75}.
\end{equation}

\end{document}